\documentclass[%
reprint,
 amsmath,
 amssymb,
 aps,
 prl,
]{revtex4-1}

\usepackage{graphicx}
\usepackage{dcolumn}
\usepackage{bm}

\usepackage{graphicx}
\usepackage[colorlinks=true,citecolor=blue]{hyperref}
\usepackage[margin=1in]{geometry}

\usepackage[caption=false]{subfig}

\usepackage[stable]{footmisc}

\usepackage{multirow}

\usepackage{physics}

\usepackage[capitalise]{cleveref}

\usepackage{xcolor}
\definecolor{bo}{rgb}{0.8, 0.33, 0.0}
\definecolor{airforceblue}{rgb}{0.36, 0.54, 0.66}
\definecolor{amethyst}{rgb}{0.6, 0.4, 0.8}
\definecolor{ao(english)}{rgb}{0.0, 0.5, 0.0}
\definecolor{armygreen}{rgb}{0.29, 0.33, 0.13}
\definecolor{spinc}{rgb}{0.0, 0.0, 0.0}

\newcommand{\hc}{ {\rm h.\,c.}}

\newcommand{\com}[2]{\left[#1,#2\right]}

\newcommand{\rhoop}{ {\hat \rho}}

\newcommand{\rhospins}{ {\hat \rho} }

\newcommand{\drhospins}{ {\dot {\hat \rho}}}

\newcommand{\mD}[2]{\mathcal{D}\left[{#1}\right] {#2} }

\newcommand{\psidk}[1]{\ket{\psi_{\rm dk}[#1]}}

\newcommand{\tauprot}{\tau_{\rm prot}}

\newcommand{\Sp}{{\color{spinc} {\hat S}_{+}}}

\newcommand{\Sm}{{\color{spinc} {\hat S }_{   -}}}

\newcommand{\Sop}{{\color{spinc} {\hat S }}}

\newcommand{\Uop}{ {\hat U }}
\newcommand{\Rop}{ {\hat R }}

\newcommand{\Sx}{{\hat S}_{x}}

\newcommand{\Sy}{{\hat S}_{y}}

\newcommand{\Sz}{{\hat S}_{z}}

\newcommand{\Sigmaop}{{\hat \Sigma}}

\newcommand{\cm}{{\hat c}}

\newcommand{\betaop}{{\hat \beta}}

\newcommand{\szz}[1]{\hat{\sigma}^{z}_{\mkern-3.0mu #1}}

\newcommand{\dg}{^\dagger}

\newcommand{\A}{{\hat A}}

\newcommand{\B}{{\hat B}}

\newcommand{\chit}{{\tilde{\chi}}}

\begin{document}


\title{Heisenberg-limited spin-squeezing via bosonic parametric driving}

\author{Peter Groszkowski$^1$, Hoi-Kwan Lau$^1$, C. Leroux$^2$, L. C. G. Govia$^3$, A. A. Clerk$^1$}
\affiliation{$^1$Pritzker School of Molecular Engineering, University of Chicago, Chicago, IL, USA \\
$^2$Institut Quantique and D\'epartement de Physique, Universit\'e de Sherbrooke, Sherbrooke J1K 2R1 QC, Canada \\
$^3$Raytheon BBN Technologies, 10 Moulton Street, Cambridge, MA, 02138, USA}

\date{\today}

\begin{abstract}
Spin-spin interactions generated by a detuned cavity are a standard mechanism for generating highly entangled spin squeezed states.  We show here how introducing a weak detuned parametric (two-photon) drive on the cavity provides a powerful means for controlling the form of the induced interactions.
Without a drive, the induced interactions cannot generate Heisenberg-limited spin squeezing, but a weak optimized drive gives rise to an ideal two-axis twist interaction and Heisenberg-limited squeezing.  Parametric driving is also advantageous in regimes limited by dissipation, and enables an alternate adiabatic scheme which can prepare optimally squeezed, Dicke-like states.  Our scheme is compatible with a number of platforms, including solid-state systems where spin ensembles are coupled to superconducting quantum circuits or mechanical modes.

\end{abstract}

\maketitle


{\it Introduction--- } 
The field of quantum sensing focuses on enhancing measurements by exploiting entanglement. Among the most studied approaches are those based on spin squeezing \cite{ma2011quantum}, where one uses an entangled state of $N$ spin 1/2 particles to reduce the imprecision of a Ramsey-type phase measurement.  While there are many approaches for generating spin squeezing (see e.g.~Refs.~\cite{kitagawa1993squeezed,
AgarwalPuri1994,schleier2010squeezing,bennett2013phonon,dalla2013dissipative,hu2017vacuum,lewis2018robust}), new methods are still of interest if they can transcend limitations of standard approaches.  The most widely studied deterministic method is based on exploiting an all-to-all Ising interaction, the so-called one-axis twist (OAT) Hamiltonian \cite{kitagawa1993squeezed}; this has been implemented in several groundbreaking experiments \cite{LerouxPRL2010,Riedel2010,GrossNature2010,HostenScience2016}.  While conceptually simple, this method cannot achieve fundamental $1/N$ Heisenberg scaling of the squeezing.  In contrast,  the so-called two-axis twist Hamiltonian (TAT) is known to achieve Heisenberg scaling \cite{kitagawa1993squeezed}, but is difficult to physically implement.    

\begin{figure}[t]
    \centering
	\includegraphics[width=0.6 \columnwidth]{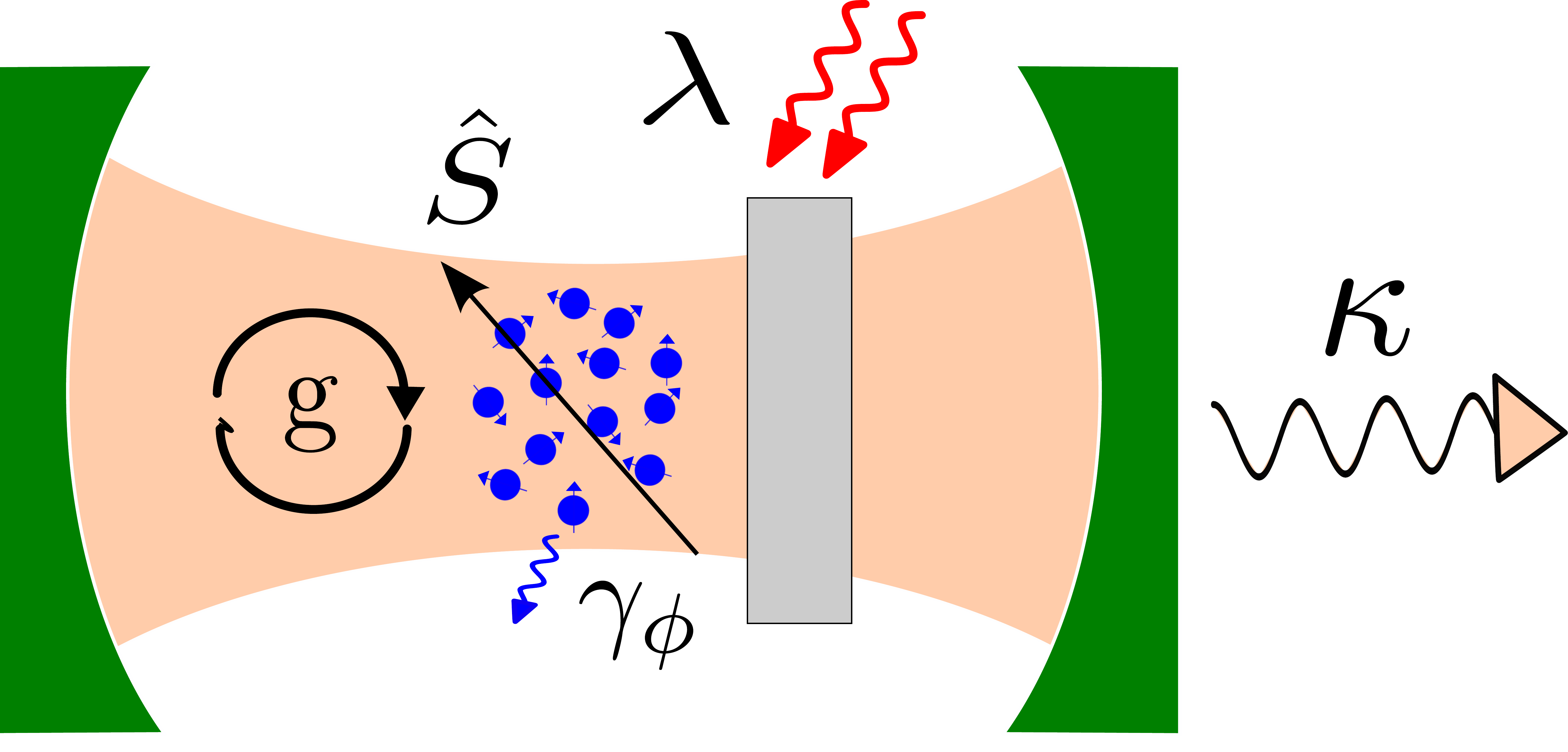} 
    \caption{A collective spin $\hat{S}$ comprised of $N$ spin $1/2$ particles is coupled to a 
    parametrically driven cavity (drive amplitude $\lambda$).  The cavity has a decay rate
    $\kappa$, and the single-spin dephasing rate is $\gamma_{\phi}$.
    \label{fig:schematic}
    }
\end{figure}

In this work, we show how adding a detuned parametric drive (PD) to the standard setup of spins coupled to a cavity (\cref{fig:schematic}) can be used to exactly implement the TAT interaction, and thus achieve Heisenberg-limited spin squeezing.  Our scheme is compatible with standard spin-echo techniques, thus giving it robustness against the effects of inhomogeneous broadening and low-frequency noise; it also outperforms standard OAT in the presence of realistic dissipation.  For stronger PD strengths, one can alternatively implement an adiabatic protocol that produces Dicke-like states which achieve the maximum possible level of spin squeezing (outperforming TAT by a factor of 2) \cite{andre2002atom,pezze2018quantum}.  Our approach could be implemented in a host of systems, including solid state spins coupled to driven mechanical modes \cite{bennett2013phonon,JayichReview2017} or driven superconducting cavities \cite{BertetPRX2017}.  

Note that our protocols differ significantly from previous ideas using 
PD for spin squeezing.
Reference~[\onlinecite{FossFeigPRL2019}] considered how PD could enhance OAT in a trapped ion setup; we consider a different basic spin-boson coupling, and demonstrate methods that go beyond OAT. Reference~[\onlinecite{NoriParametric2019}] considered how PD in consort with strong cavity frequency modulation could realize dissipative spin squeezing \cite{AgarwalPuri1990} through a higher-order process.  Our approaches in contrast require no frequency modulation, and have far more favourable cooperativity requirements as they do not involve higher-order processes.

{\it Model--- }  
We consider $N$ two-level systems (splitting frequency $\omega_s$) coupled via a standard Tavis-Cummings interaction (strength $g$) to a bosonic mode subject to a parametric (i.e.~two-photon) drive at frequency $2 \omega_p$:
\begin{align}
    \hat{H}_{\rm lab} &=   \omega_{c}  \cm \dg \cm + \omega_{s} \Sz 
    + \left( g \cm  \Sp + \frac{\lambda}{2}  e^{i 2 \omega_{p}t}  \cm^{2}    + \hc \right),
    \label{eq:Hlab}
\end{align}
where we introduce collective spin operators
$\hat{S}_{\pm}=\Sx \pm i \Sy$ and
$\hat{S}_{k} =\frac{1}{2} \sum_{j} \hat{\sigma}^{k}_{(j)}$ for $k \in \{x,y,z\}$, with $\hat{\sigma}^{k}_{(j)}$ denoting a standard Pauli operator acting on the $j$th spin. 
The parametric drive will give us a powerful means for controlling the form of the cavity mediated spin-spin interactions.
We next move to a rotating frame (for both spins and cavity) in which the Hamiltonian is time-independent:
\begin{align}
    \hat{H}_{\rm rot} &=    
    	\Delta_c \cm \dg \cm  + \Delta_s \hat{S}_z +
    	\left( g \cm \dg \Sm + \frac{1}{2}  \lambda   \cm^{2} + \hc \right).
    \label{eq:Hrot}
\end{align}
Here $\Delta_{c/s} \equiv \omega_{c/s} - \omega_p$ are the respective detunings of the cavity and spins from the parametric drive.  

Without loss of generality, we take the parametric drive amplitude $\lambda$ to be real and positive, and consider the regime $|\Delta_c| \geq \lambda$, ensuring a stable system.  We can then diagonalize the cavity Hamiltonian in terms of a Bogoliubov mode $\betaop \equiv  \cosh r \cm + \sinh r \cm \dg$, where the parameter $r$ satisfies 
$ \tanh 2r = \lambda / \Delta_c$.  Defining $E_\beta \equiv \sqrt{ \Delta_c^2 - \lambda^2}$, this yields
\begin{align}
    \hat{H}_{\rm sq} &=  E_\beta \betaop \dg \betaop  + \Delta_s \hat{S}_z
    	+ g \left( \betaop \dg  \Sigmaop + \hc \right).
    \label{eq:Hdiag}
\end{align}
where the spin Bogoliubov mode is defined as 
$\Sigmaop \equiv  \cosh r \Sm  - \sinh r \Sp $. 

We next consider the case where $\sqrt{N} g \ll E_{\beta}$, and where the parametric drive is almost resonant with the spins, such that $\Delta_s \sim g^2 / E_{\rm \beta} \ll E_{\rm \beta}$.  In this case, we can eliminate the cavity-spin interaction to leading order using
a standard Schreiffer-Wolff transformation (see \cite{supp}); this is analogous to standard derivations of cavity-mediated OAT \cite{AgarwalPRA1997,bennett2013phonon}.  Retaining terms to order $g^2$, we obtain an effective interacting spin Hamiltonian:
\begin{align}
    \hat{H}_{\rm eff} & \simeq 
        E_\beta \betaop \dg \betaop + \Delta_s \hat{S}_z
    	- \chi \Sigmaop \dg \Sigmaop   - \chi  \Sz \betaop \dg \betaop. 
    \label{eq:Heff}
\end{align}
with $\chi \equiv g^2 / E_{\beta}$.  
Superficially, this is identical to the Hamiltonian for cavity-mediated OAT, except the spin lowering operator has been replaced by $\Sigmaop$, the spin Bogoliubov operator.  As we now show, this has dramatic consequences.


\begin{figure}[t]
    \includegraphics[width=0.95 \columnwidth]{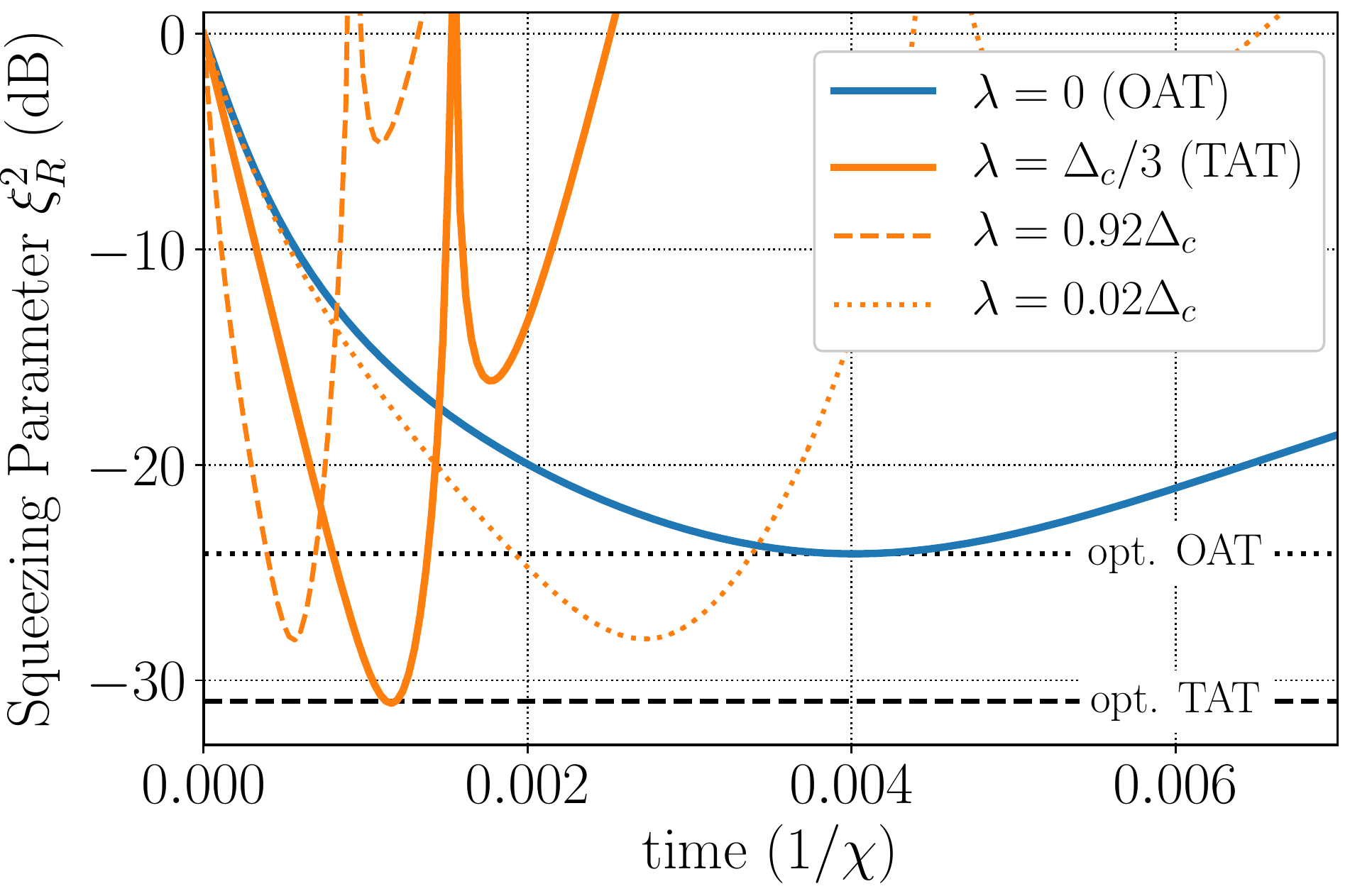}
    \vspace{-0.2cm}
    \caption{
        Dissipation-free evolution of spin squeezing under the spin-spin interaction in \cref{eq:effSpins1} for different parametric drive amplitudes ($\tilde{\Delta}=0$, $N=5000$, spins initially polarized along $x$). The solid blue (orange) curve corresponds to the OAT (TAT) evolution with the parametric drive amplitude $\lambda=0$ ($\lambda=\Delta_{c}/3$). The dotted (dashed) orange curve shows results for non-optimal amplitude of $\lambda=0.02\Delta_c$ ($\lambda=0.92\Delta_c$). Even non-ideal choices of $\lambda$ lead to performance that surpasses OAT.  Horizontal lines indicate the optimal squeezing for OAT and TAT.  We assume that a Hanh-echo has been performed to cancel the effects of the dispersive interaction.  In the absence of this echo, the optimal squeezing for $\lambda = \Delta_c/3$ is degraded by less than $1$ dB.    }
    \label{fig:ITATCoherent}
\end{figure}

{\it Induced two-axis twisting (ITAT)--- }  We first ignore the last dispersive coupling term in Eq.~(\ref{eq:Heff}).  In this case the spins and cavity are decoupled, and the spin-only terms in Eq.~(\ref{eq:Heff}) describe an unusual kind of cavity-mediated spin-spin interaction.  Expanding these terms out, and defining $\tilde{\chi} = \chi \cosh 2 r$, $\tilde{\Delta} = \Delta_s - \chi$, we have:
\begin{align}
    \hat{H}_{\rm s} &= 
        \tilde{\Delta} \hat{S}_z
    - \tilde{\chi} \Bigg[ 
    	 \left( \hat{S}_{\rm tot}^2 - \Sz^{2} \right ) 
     - \tanh(2r) \left( \Sx^{2} - \Sy^{2} \right) \Bigg],
    \label{eq:effSpins1}
\end{align}
with $\hat{S}_{\rm tot}^{2} = \hat{S}_{x}^{2} + \hat{S}_{y}^{2} + \hat{S}_{z}^{2} $.
Without a parametric drive (i.e.~$r=0$) we have a standard cavity-induced OAT Hamiltonian \cite{AgarwalPRA1997,bennett2013phonon}.  For non-zero $r$, the new interaction terms have the form of the TAT Hamiltonian introduced in Ref.~\cite{kitagawa1993squeezed}; these terms on their own are capable of generating spin squeezing with Heisenberg-limited scaling, something that is impossible with an OAT Hamiltonian.  

At first glance, it seems like our scheme can never realize a pure TAT interaction, both because the OAT-like terms will always dominate (as $\tanh 2 r \leq 1$), and because of the spurious linear-in-$\Sz$ term.  This pessimism is unfounded.  First, the unwanted linear term can be eliminated by simply tuning the spin detuning to $\Delta_s = \chi$; this could be done, e.g., by just slightly shifting the parametric drive frequency.  Second, if we also tune the parametric drive amplitude so that 
$\lambda = \Delta_{c} / 3$, we have $\tanh 2r \rightarrow \tanh 2r_0 = 1/3$, and the resulting Hamiltonian can be written:
\begin{align}
    \hat{H}_{s} & \rightarrow
    - \tilde{\chi}  \Big[ 
    	 \left( \hat{S}_{\rm tot}^2 - \Sz^{2} \right ) 
	 - \frac{1}{3} \left( \Sx^{2} - \Sy^{2} \right)  \Big]
	\nonumber \\
	& = 
   - \frac{2}{3} \tilde{\chi} \Big[
   	 \hat{S}_{\rm tot}^2  
    	  -\Sz^{2}   
	 +  \Sy^{2}   \Big].
    \label{eq:ITATH}
\end{align}
Since $\hat{S}_{\rm tot}$ is a constant of motion for $\hat{H}_s$, the effective dynamics of $\hat{H}_{s}$ are equivalent to the desired two-axis twist Hamiltonian.  

Eq~(\ref{eq:ITATH}) is a central result of our work, and represents one the simplest mechanisms we know for implementing the TAT Hamiltonian.  Previous proposals for realizing TAT 
tend to be experimentally demanding.  They either require carefully tailored bang-bang control of the spin ensemble \cite{Cappellaro2009,You2011TAT},  multiple drive lasers, atomic levels and cavity transitions \cite{borregaard2017one}, or very weak higher-order interaction processes \cite{NoriTATPreprint}.  In contrast, our scheme utilizes a standard Tavis-Cummings coupling, and does not require an elaborate pulsed driving of the spin system.  It also requires only a modest-amplitude parametric drive (far from any regime of instability).      Note one could alternatively tune $\lambda = - \Delta_{c} / 3$; in this case an equivalent TAT Hamiltonian in the $z-x$ plane is generated.  By tuning the parametric drive amplitude, one can also realize other kinds of spin-spin interactions, including an OAT Hamiltonian along the $y$ axis, and a ``twist-and-turn"  Hamiltonian \cite{Strobel2014,Oberthaler2015} (see \cite{supp}).

We now return to the issue of the dispersive interaction in Eq.~(\ref{eq:Heff}). In the absence of dissipation, $\hat{\beta}^\dagger \hat{\beta}$ is a conserved quantity.  
Further, assuming the cavity starts in a vacuum state, the $\betaop$ mode starts in a squeezed state characterized by $r_{0}$, and thus has a small but non-zero population.
Hence, the small mean value $\langle \betaop \dg \betaop \rangle = \sinh^2 r_0 \simeq 0.03$ can be easily cancelled by slightly shifting the spin-drive detuning to $\Delta_s = \chi (1 + \sinh^2 r_0)$. The remaining static fluctuations of the Bogoliubov-mode number operator have a dephasing effect, which is also insignificant due to the smallness of the required parametric drive amplitude 
(i.e.~$\langle (\hat{\beta}^\dagger \hat{\beta})^2 \rangle - \langle (\hat{\beta}^\dagger \hat{\beta}) \rangle^{2} \simeq 0.06$). 
They have a negligible effect on the optimal squeezing 
(numerical simulations show that at $N=5000$, the change in $\xi_{R}^{2}$ is much smaller than $1\,$dB), and furthermore, their effects can be {\it completely} cancelled by performing a single Hahn-echo pulse half-way through the evolution period (corresponding to a $\pi$ pulse about, e.g., the $x$ axis).  This highlights another key advantage of our scheme:  like the standard cavity-based OAT \cite{bennett2013phonon}, it is fully compatible with widely used spin-echo techniques for suppressing the effects of inhomogeneous broadening and low-frequency dephasing. This is of particular importance in potential solid-state implementations.        

As is standard, we quantify the amount of useful spin squeezing using the Ramsey spin squeezing parameter \cite{wineland1994squeezed}:
\begin{align}\label{eq:wp}
	\xi_R^2 \equiv 
		N  \langle \Delta \hat{S}_{\perp}^2 \rangle / 
		\left \langle \vec{\hat{S}} \right \rangle^2, 
\end{align}
where $\Delta \hat{S}_{\perp}^{2}$ is the minimum variance in a direction perpendicular to the direction of the mean of the collective spin.  For our induced TAT Hamiltonian, we start with an initial product state where all spins are polarized along the $x$ direction.  
As shown in \cref{fig:ITATCoherent}, if we use a Hahn echo to cancel the effects of the dispersive coupling, the induced TAT Hamiltonian (in the absence of dissipation) generates spin squeezing at an optimal time,
 that scales as $\xi^2_R \sim 4/N$.  This represents Heisenberg-limited scaling, something that is impossible with a standard OAT protocol (i.e.~our setup with zero parametric drive), where 
 $\xi^2_R \sim 1/N^{2/3}$ at best. 
 Figure~\ref{fig:ITATCoherent} also shows that our protocol is robust against variations in the parametric drive amplitude; even when $\lambda$ is far away from its optimal value of $\Delta_{c}/3$ the performance is superior to OAT.


{\it Impact of dissipation--- } 
It is also crucial to understand the ITAT scheme in the presence of dissipation.  As discussed, standard spin-echo pulses are compatible with our scheme, and hence can be used to suppress the impact of inhomogeneous broadening and low-frequency dephasing noise.  For the remaining dissipative processes, we assume each spin is dephased by a Markovian bath (rate $\gamma_{\phi}$) and that the cavity has a energy damping rate $\kappa$ due to coupling to a zero-temperature environment.  The dissipative dynamics of our system is then described by
  \begin{align}
    \drhospins &=  -i [\hat{H}_{s}, \rhospins] 
    +  \mD{ \sqrt{\Gamma} \hat{z}[r] }{\rhospins} 
    + \frac{\gamma_{\phi}}{2} \sum_{k=1}^N \mD{\szz{(k)}}{\rhospins},
    \label{eq:meEff}
\end{align}
where 
$\mathcal{D}[z]\hat{\rho} = \hat{z} \hat{\rho} \hat{z}^\dagger - \{ \hat{z}^\dagger \hat{z}, \hat{\rho} \}/2$ is the standard Linblad dissipative superoperator.  $\Gamma = \kappa \chi / E_{\beta}$ is the rate associated with cavity-induced spin dissipation; the jump operator describing this process is
\begin{align}
	\hat{z}[r]   
		= e^{-2 r} \hat{S}_x - i e^{2 r} \hat{S}_y.
 \end{align}
 For large parametric drives (e.g.~as used in the scheme of Ref.~\cite{FossFeigPRL2019}), the drive causes strong amplification of the cavity-induced dissipation, potentially nullifying any advantage.  In contrast, our scheme only requires a small parametric drive 
 (i.e.~$e^{2 r_0} = \sqrt{2}$), leading to minimal amplification of dissipation.

\begin{figure}[t]
    \includegraphics[width=0.9 \columnwidth]{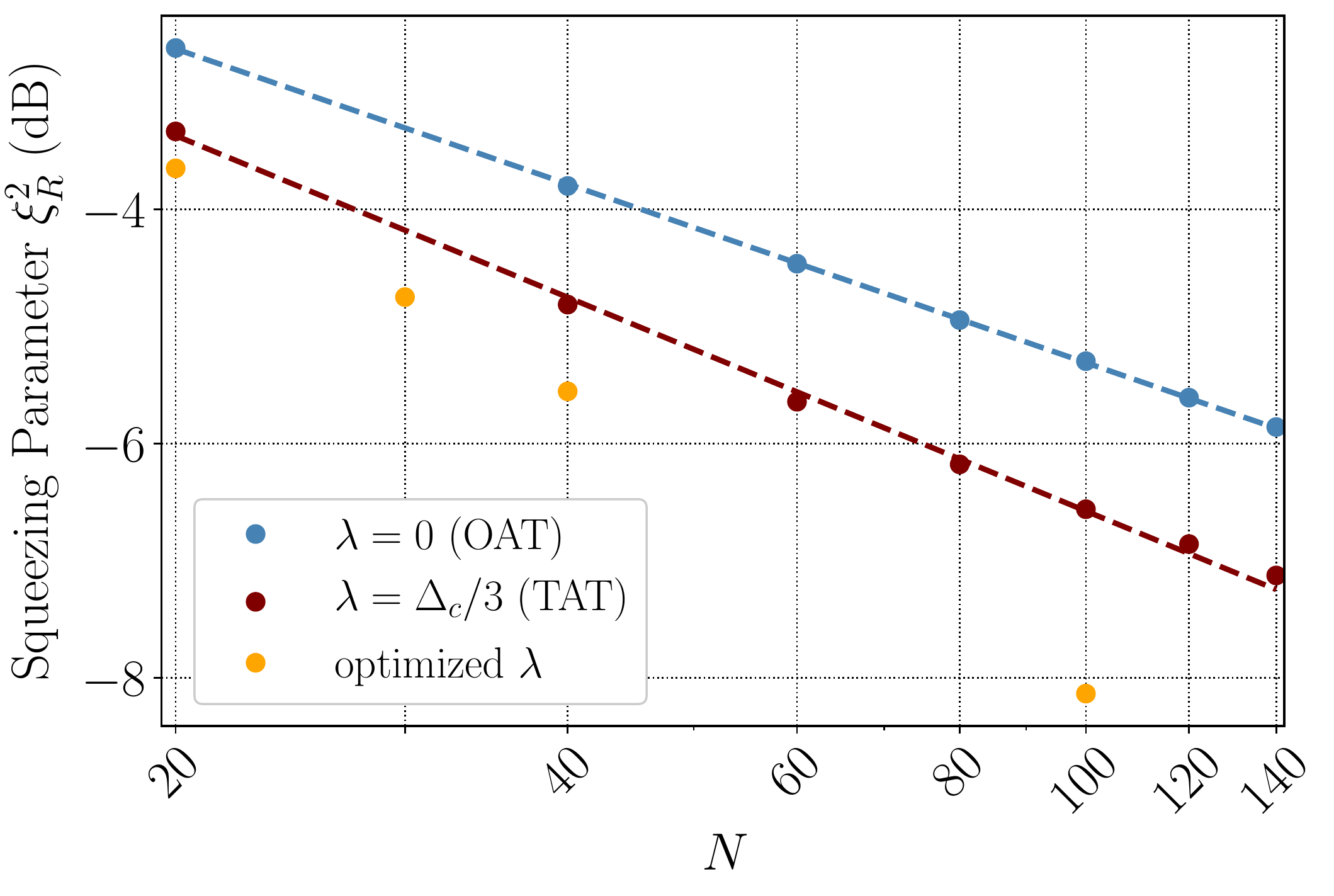}
    \vspace{-0.2cm}
    \caption{ 
        Optimized squeezing parameter $\xi_{R}^{2}$ in the presence of dissipation versus number of spins. The dots represent the effective master equation (\cref{eq:meEff}) simulation data for OAT (blue) corresponding to $\lambda=0$, ITAT (maroon) corresponding to $\lambda=\Delta_c/3$ and finally a case where the drive strength $\lambda$ is itself (approximately) optimized (orange), with values $\lambda=\alpha \Delta_{c}$, where $\alpha=0.70,\,0.74,\,0.76,\,086$ for $N=20,\,30,\,40,\,100$ respectively. In all cases we take $\kappa=10g$, $\gamma_{\phi}=0.02g$ and optimize over $E_{\beta}$ and protocol time. The dashed lines depict corresponding numerical fits to $a\mathcal{C}^{-b}$, with $3.2\mathcal{C}^{-0.4}$ (blue), $3.8\mathcal{C}^{-0.5}$ (maroon).
    }
    \label{fig:ITATDissipation}
\end{figure}

 Shown in \cref{fig:ITATDissipation} are results from numerical simulations of the full master equation \cite{johansson2013qutip,shammah2018open}, depicting optimal spin squeezing versus $N$ (with $E_\beta$ optimized for each $N$).  We pick
 parameters such that $\kappa \gg \gamma_\phi$ and the collective cooperativity $\mathcal{C} \equiv N g^2 / (\kappa \gamma_\phi)$ is $5$ for $N=1$, and always evolve starting with spins fully polarized in the $x$ direction.  Even with dissipation, a parametric drive corresponding to $r = r_0$ (i.e.~ITAT) appreciably improves performance for all values of $N$ over the undriven ($r=0$, OAT) case. Our results are also consistent with an approximate $\xi^2_R \sim 1 / \sqrt{\mathcal{C}}$ scaling, as would be expected from a standard linearized treatment of our system (see \cite{supp}). We also consider optimizing the value of $r$ (i.e.~parametric drive strength) for each $N$.  We find that these optimized $r$ values (orange points in ~\cref{fig:ITATDissipation}), are always larger than the value $r_0$ that would yield the TAT Hamiltonian.  
At a heuristic level, increasing $r$ increases the initial rate at which squeezing is produced, something that is likely advantageous in the presence of dissipation.

{\it Adiabatic preparation of optimally squeezed states--- } While the TAT Hamiltonian is able to produce Heisenberg-limited spin squeezing, it is well known that states exist which are squeezed by an {\it additional} factor of 2 \cite{andre2002atom,pezze2018quantum}.  Such states are infinitesimally close to so-called ``Dicke states'':  collective spin eigenstates that have a maximal $\hat{S}_{\rm tot}^2$ and are also annihilated by, e.g.~$\hat{S}_z$.  As we now show, by making our parametric drive time-dependent, our setup can also produce such states.
 
To see how this works, note that for even $N$ the spin Bogoliubov operator $\hat{\Sigma}[r]$ has a unique state in its kernel, $\ket{\psi_{\rm dk}[r]}$ 
\cite{AgarwalPuri1990,AgarwalPuri1994,dalla2013dissipative}. This state exhibits optimal spin-squeezing properities, with $\xi^2_R \rightarrow 2 / N$ in the large-$r$ limit.
Furthermore, it can be naturally produced by driving a spin ensemble with squeezed light \cite{AgarwalPuri1990,AgarwalPuri1994}, and can also be stabilized using dissipative protocols involving multi-level atoms and engineered Raman processes \cite{dalla2013dissipative}.

\begin{figure}[t]
    \includegraphics[width=0.95 \columnwidth]{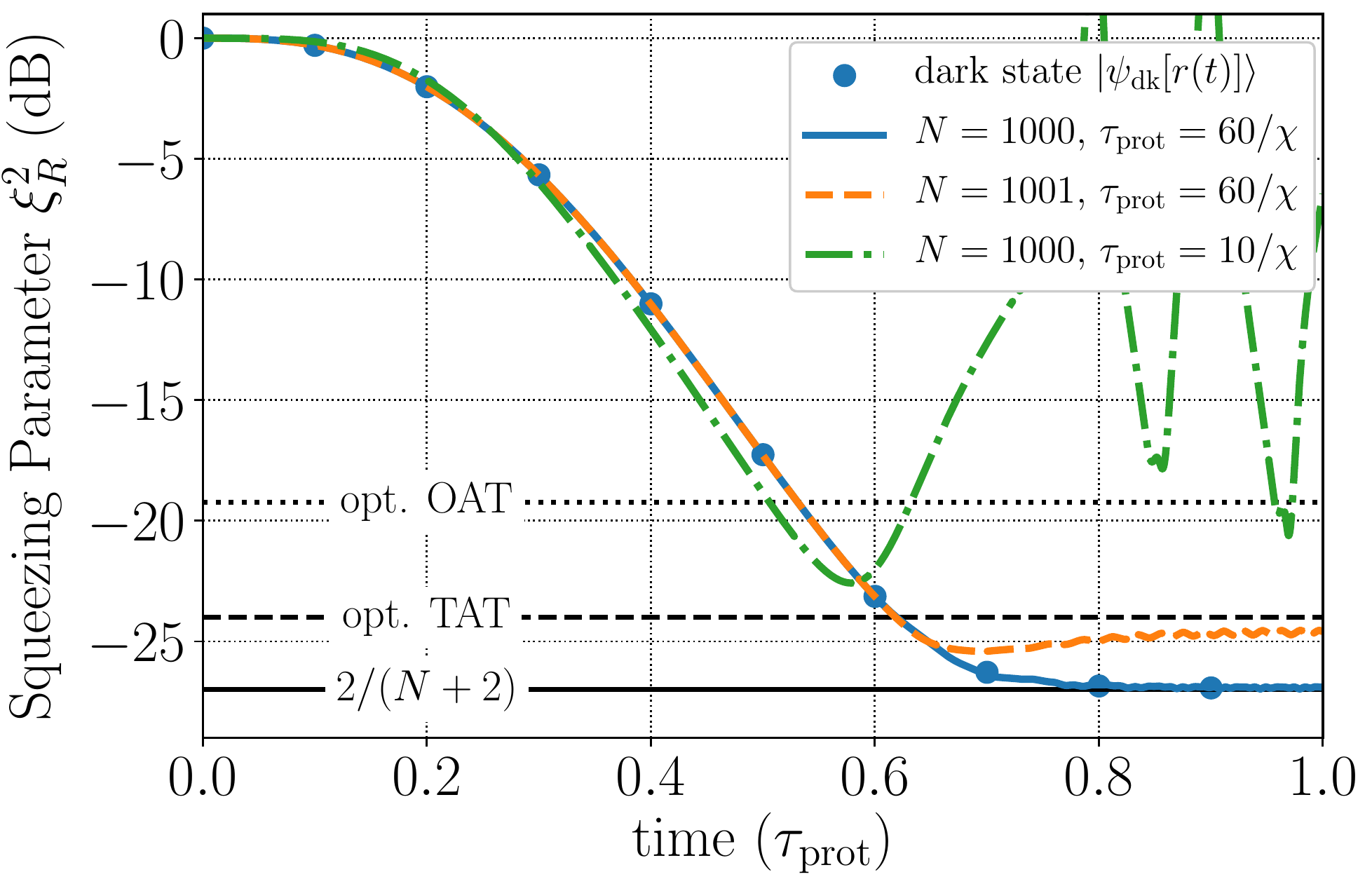}
    \vspace{-0.2cm}
    \caption{Squeezing parameter $\xi_{R}^{2}$ versus time for the adiabatic scheme, for different protocol times:   
    $\tau_{\rm prot}=60/\chi$ (blue, solid), $\tau_{\rm prot}=10/\chi$ (green, dash-dotted). In both cases $N=1000$. The blue dots correspond to the performance one would obtain from an ideal dark state $\psidk{r(t)}$. 
        The orange dashed curve depicts the evolution with $\tau_{\rm proc}=60/\chi$, but with an odd number of spins; $N=1001$. 
    }
    \label{fig:Adiabatic}
\end{figure}

Our setup provides an alternate, fully coherent method for generating such states.  For even $N$, $\ket{\psi_{\rm dk}[r]}$ is the unique zero-energy ground state of the drive-modified spin-spin interaction in Eq.~(\ref{eq:Heff}); all other higher-lying states are separated by a gap.  
  As discussed in \cite{supp}, by making the complex parametric drive amplitude $\lambda$, frequency $\omega_p$ and spin Larmor frequency $\omega_s$ all time-dependent, we obtain a Hamiltonian with the same form as \cref{eq:Heff}, except with a time-dependent squeezing parameter $r(t)$.
  This Hamiltonian always has an instantaneous zero-energy eigenstate $\ket{\psi_{\rm dk}[r(t)]}$.  Our protocol thus consists of starting with $\lambda(0)=r(0)=0$, with an initial state having of all the spins polarized along the $z$ axis (i.e.~$\psidk{r=0} = \ket{N/2, -N/2}$).  We then slowly ramp up $r(t)$ from $0$ to $r_f$ by appropriately varying $\lambda(t)$, $\omega_{p}(t)$ and $\omega_{s}(t)$ (see \cite{supp}).  The adiabatic theorem then implies that the system evolves from its initial product form to the highly entangled state $\ket{\psi_{\rm dk}[r_f]}$. One can show analytically that for large $r_f$, $\ket{\psi_{\rm dk}}$ exhibits spin squeezing with $\xi^2_R \sim 2/N$ \cite{AgarwalPuri1990,AgarwalPuri1994}.
  Adiabaticity requires a total evolution time $\tauprot$ that is much longer than the relevant inverse gap, which in our case scales as $N g^2 / E_{\beta}$.  

  Shown in \cref{fig:Adiabatic} are numerical results for the time-evolution of the squeezing under this adiabatic protocol for $N=1000$ and for different total protocol times.  We take the time-dependence of $r(t)$ to smoothly increase from 0 to 4 during the evolution (see \cite{supp}). Even for faster evolution times where non-adiabatic errors are prevalent, large amounts of spin squeezing are produced, and performance can still surpass that of standard OAT.  
In practice, the dark state
 $\ket{\psi_{\rm dk}[r(t)]}$ only exists for even $N$. Figure~\ref{fig:Adiabatic} shows that our protocol nonetheless produces considerable squeezing even when $N$ is odd.  


{\it Conclusions--- } 
We have explored how parametrically driving a cavity coupled to a spin ensemble can be used to optimize the generation of highly squeezed spin states.  An optimally detuned parametric drive allows a direct realization of the ideal TAT spin-squeezing Hamiltonian, enabling Heisenberg-limited scaling.  This protocol also significantly improved performance over the undriven system in regimes limited by dissipation.  We also described an alternate protocol using a time-dependent parametric drive, which adiabatically produced optimally spin-squeezed states which approach Dicke states. 

P.G., H-K.L. and A.A.C. acknowledge support by the DARPA DRINQS program (agreement D18AC00014).

\bibliography{library}





\newpage 
\begin{appendix}
\clearpage
\thispagestyle{empty}
\onecolumngrid
\begin{center}
\textbf{\large Supplemental Material: Heisenberg-limited spin-squeezing via bosonic parametric driving}
\end{center}

\setcounter{equation}{0}
\setcounter{figure}{0}
\setcounter{table}{0}
\setcounter{page}{1}
\makeatletter
\renewcommand{\theequation}{S\arabic{equation}}
\renewcommand{\thefigure}{S\arabic{figure}}
\renewcommand{\bibnumfmt}[1]{[S#1]}
\renewcommand{\citenumfont}[1]{S#1}

\section{Effective Induced TAT System Hamiltonian and Master Equation}
\label{app:effectiveHandME}

In this section, we present a self-contained discussion with additional details related to obtaining an effective system Hamiltonian as well as the corresponding master equation that in the main text we use to induce TAT squeezing dynamics of the spins. We start by considering a parametrically driven cavity represented by an annihilation operator $\cm$, that is coupled to a collective spin $\Sop$, with a Hamiltonian (taking $\hbar=1$)
\begin{align}
    \hat{H}_{\rm lab} &= \omega_{c} \cm \dg \cm + \omega_{s} \Sz + \left( g \cm \dg \Sm + \frac{\lambda}{2} \cm \dg \cm \dg e^{-i 2\omega_{p}t} + \hc \right),
    \label{eq:appH1}
\end{align}
where $\omega_{c}$ is cavity frequency, $\omega_{s}$ the collective spin frequency, $g$ the single-spin cavity-spin coupling strength, $\omega_{p}$ half of the parametric drive frequency, and finally $\lambda$ the parametric drive amplitude. Without a loss of generality, we assume $\lambda$ is real and positive.
Furthermore, we envision that our cavity-spin system is subjected to two noise channels: cavity decay at rate $\kappa$, as well as local spin dephasing (uniform among all spins), characterized by the rate $\gamma_{\phi}$. This leads to a master equation 
\begin{align}
    \dot\rhoop &=  -i [\hat{H}_{\rm lab}, \rhoop] + \kappa \mD{\cm}{\rhoop} 
    + \frac{\gamma_{\phi}}{2} \sum_{k} \mD{\szz{(k)}}{\rhoop},
    \label{eq:appMe1}
\end{align}
where $\mD{\A}{\rhoop} = \A \rho \A \dg - \left( \A \dg \A \rhoop + \rhoop \A \dg \A \right)/2 $, for any operator $\A$.

{\it Effective Hamiltonian--- } 
In order to arrive at an effective description of the above setup, we begin with the Hamiltonian from \cref{eq:appH1}, and go into a rotating frame dictated by $\omega_{p}$. Introducing $\Delta_{c}=\omega_{c} - \omega_{p}$ and $\Delta_{s}=\omega_{s} - \omega_{p}$ lets us write
\begin{align}
    \hat{H}_{\rm rf} &= \Delta_{c} \cm \dg \cm + \Delta_{s} \Sz   + g \left( \cm \dg \Sm + \hc \right) + \frac{\lambda}{2} \left( \cm \dg \cm \dg + \hc \right).
    \label{eq:appH2}
\end{align}
Next, using 
\begin{align}
    \Uop_{\rm sq} &=  e^{-\frac{r}{2} \left(  \cm^{2} - (\cm \dg)^{2} \right)}, 
    \label{eq:Usq}
\end{align}
we diagonalize the first and last terms of \cref{eq:appH2}. Choosing $r$ such that $\tanh(2r) = \lambda/\Delta$, leads to
\begin{align}
    \hat{H}_{\rm sq} &=  \Uop_{\rm sq}  \hat{H}_{\rm rf} \Uop_{\rm sq} \dg  \nonumber \\
    &=  E_{\beta} \betaop \dg \betaop  +  \Delta_{s} \Sz + \left(g \betaop \dg \Sigmaop + \hc \right),
    \label{eq:appH3}
\end{align}
with $E_{\beta} = \Delta_{c} \sech(2r) = \sqrt{\Delta^{2}  - \lambda^{2}}$, and where for convenience we have introduced a Bogoliubov-like spin-mode
\begin{align}
    \Sigmaop &=  \cosh r \Sm  - \sinh r \Sp,
    \label{eq:sigmaop1}
\end{align}
as well as a Bogoliubov cavity mode 
\begin{align}
    \betaop &=  \cosh r \cm + \sinh r \cm \dg. 
    \label{eq:betaop1}
\end{align}

{\it Cavity Elimination--- } In order to simplify \cref{eq:appH3} further, we eliminate the cavity from the problem, using a Schrieffer-Wolff transformation \cite{schrieffer1966relation,bennett2013phonon}. Assuming $ \sqrt{N} g \ll E_{\beta} $ and using a generator  
\begin{align}
    \Rop &= \frac{g}{E_{\beta}} \left(\betaop \dg  \Sigmaop - \hc \right), 
    \label{eq:appRdef}
\end{align}
we calculate $\tilde{H}_{\rm eff} = e^{\Rop} \hat{H}_{\rm sq} e^{-\Rop}$. Using the fact that $\com{\betaop}{\betaop \dg}=1$, $\com{\Sigmaop \dg}{\Sigmaop} = 2\Sz$ as well as $\com{\Sz}{\Sigmaop}=\Sigmaop$, and keeping only the leading order terms in $g / E_{\beta}$, we write
\begin{align}
    \hat{H}_{\rm eff} &\simeq   E_\beta \betaop \dg \betaop  + \Delta_{s} \Sz - \chi \Sigmaop \dg \Sigmaop   - \chi  \Sz \betaop \dg \betaop, 
    \label{eq:appHeff}
\end{align}
where for convenience we have defined $\chi = g^{2} /  E_{\beta}$, and assumed that $\Delta_s \sim g^2 / E_{\rm \beta} \ll E_{\rm \beta}$.
In \cref{eq:appHeff} the cavity-spin degrees of freedom now only couple dispersively. By assuming that this dispersive interaction can be (on average) eliminated through a dynamical decoupling protocol (by appropriate rotations along a suitably chosen axis in the $x$-$y$ plane) or by choosing appropriate spin detuning (see below), we can concentrate only on the spin degree of freedom, and write 
\begin{align}
    \hat{H}_{s} &= -\chi \Sigmaop \dg \Sigmaop  \nonumber  + \Delta_{s} \Sz \\
     &= 
    - \chit \Big[ 
    	 \left( \hat{S}_{\rm tot}^2 - \Sz^{2} \right ) 
	 - \tanh(2r) \left( \Sx^{2} - \Sy^{2} \right) \Big]  
     + \tilde{\Delta} \Sz,  
    \label{eq:appEffH2}
\end{align}
with $\tilde{\chi} = \chi \cosh 2 r$ and $\tilde{\Delta} = \Delta_s - \chi$. We stress that the last term can be easily tuned by varying $\Delta_{s}$, which can be useful for both generating the ideal Hamiltonian, but also for removing the mean effects of the dispersive cavity-spin interaction in \cref{eq:appHeff} (see main text).

{\it Induced OAT Dynamics---}
Let us for the moment assume $\Delta_{s}=0 \rightarrow \tilde{\Delta}=-\chi$.
In the limit where $r=0$ ($\lambda=0$), it is clear that \cref{eq:appEffH2} reduces to a Hamiltonian of a standard OAT protocol along the $z$ axis, namely
\begin{align}
    \hat{H}_{s} &\rightarrow 
    - \chi
    \left( \hat{S}_{\rm tot}^2 - \Sz^{2} + \Sz \right).
    \label{eq:appOAT1}
\end{align}
Here our results simply resemble the standard method of generating OAT evolution using a dispersively coupled cavity-spin system (for example as in \cite{bennett2013phonon}). 
We also note that the term proportional to $\Sz$ can be eliminated by choosing $\tilde{\Delta}=0$, although its presence has no impact on the amount of squeezing \cref{eq:appOAT1} can produce. 

We can also obtain an approximate OAT Hamiltonian along the $y$ axis in the limit where $r \rightarrow \infty$ ($\lambda \rightarrow \Delta_{c}$). Noting that when $r \rightarrow \infty$, we have $\cosh 2r \approx \sinh 2r \approx e^{2r}/2$, which lets us rewrite \cref{eq:appEffH2} (still with $\Delta_{s}=0$) as
\begin{align}
    \hat{H}_{s} &\rightarrow - \frac{1}{2} \chi e^{2r} \left[ \left( \Sop_{\rm tot}^{2} - \Sz^{2} - \Sx^{2} + \Sy^{2} \right) + e^{-2r} \Sz \right] \nonumber \\
&= - \chi e^{2r}\left( \Sy^{2} + e^{-2r} \Sz \right) \nonumber \\
& \approx - \chi e^{2r}  \Sy^{2}. 
    \label{eq:appOAT2}
\end{align}
In the above expression, the effective spin-spin interaction strength is now enhanced (i.e.~$\chi \rightarrow \chi e^{2r}$), and points along a different axis than the $r=0$ limit shown in \cref{eq:appOAT1}.  In the presence of distinct $T_1$ and $T_2$ spin relaxation processes, different choices of the OAT axis will lead to different amounts of squeezing.  Further, note that if we keep the spin detuning $\Delta_s$, then there will be an additional $\hat{S}_z$ term in \cref{eq:appOAT2}. This then realizes the twist-and-turn Hamiltonian studied by many authors \cite{Strobel2014,Oberthaler2015}.

{\it Induced TAT Dynamics---}
Going back to \cref{eq:appEffH2}, we consider another, far more interesting scenario by choosing values of $r$ that satisfy $\tanh 2r = \pm 1/3$.  
Along with setting $\tilde{\Delta}=0$, such a condition lets us simplify \cref{eq:appEffH2} to 
\begin{align}
    \hat{H}_{s} & \rightarrow  
   - \frac{2}{3} \tilde{\chi} \Big[
    \Sop^{2}_{\rm tot}
    	  -\Sz^{2}   
      +  \Sy^{2}   \Big],
    \label{eq:appTAT1}
\end{align}
in the case of $\tanh 2r = 1/3$ ($\lambda=\Delta_c/3$), and a similar expression, with $\Sx \rightarrow \Sy$, in the case of $\tanh 2r = - 1/3$ ($\lambda=-\Delta_c/3$). 
Clearly \cref{eq:appTAT1} corresponds to a TAT Hamiltonian, which can lead to optimal $1/N$ scaling of the $\xi_{R}^{2}$ parameter. 

{\it Dissipation and Effective Master Equation--- }
In this section, we discuss how the above treatment changes the master equation shown in \cref{eq:appMe1}. 
We start by applying the Schrieffer-Wolff transformation generated by $\Rop$, to the cavity dissipator proportional to $\kappa$ from \cref{eq:appMe1}.
Keeping the leading order correction in $g/E_{\beta}$ and using \cref{eq:betaop1}, we have
\begin{align}
    e^{\Rop}  \cm e^{-\Rop} & \approx   \left( \cosh r \betaop  - \sinh r \betaop \dg \right) - \frac{g}{E_{\beta}} \left( \cosh 2r \Sm - \sinh 2r \Sp  \right).
    \label{eq:cmSW}
\end{align}
Then, noting that for any operators $\A$ and $\B$
\begin{align}
    \mD{\A + \B}{\rhoop} &= \mD{\A}{\rhoop} + \mD{\B}{\rhoop}  + \mD{\A, \B}{\rhoop} + \mD{\B, \A}{\rhoop},
\end{align}
with
\begin{align}
    \mD{\A, \B}{\rhoop} &= \A \rhoop \B\dg - \frac{1}{2} \{ \A\dg \B, \rhoop \} = \A \rhoop \B\dg - \frac{1}{2} \left( \A\dg \B \rhoop +  \rhoop \A\dg \B \right),
\end{align}
lets us approximate
\begin{align}
    \kappa \mD{\cm}{\rhoop} &\approx \kappa \mD{\cosh r \betaop  - \sinh r \betaop \dg }{\rhoop} 
    + \Gamma \mD{ \cosh 2r \Sm - \sinh 2r \Sp }{\rhoop},
    \label{eq:cDisApprox1}
\end{align}
where we have defined the effective dissipative rate of the collective spins degree of freedom term as $\Gamma = \kappa \chi / E_{\beta} $, as well as dropped the fast oscillating cross-terms (proportional to $\mD{\betaop, \Sp}{\rhoop}$, etc.). 
This lead to the effective master equation for the spins as shown in the main text. 

\section{Adiabatic evolution protocol}
\label{app:effectiveHforDicke}

{\it Effective Hamiltonian--- } 
As discussed in the main text, our scheme can also be used to adiabatically prepare optimally spin-squeezed states.  This requires engineering a Hamiltonian with the same form as \cref{eq:appH3}, except where the $r$ parameter is time-dependent. Similarly to \cref{eq:appH1}, we start with 
\begin{align}
    \hat{H}_{\rm lab}(t) &= \omega_{c} \cm \dg \cm + \omega_{s}(t) \Sz + g \left( \cm \dg \Sm + \cm \Sp \right) + \left(  \frac{\lambda(t)}{2} \cm^{2}  e^{i \varphi_{p}(t)}  + \frac{\lambda(t)^{*}}{2} ( \cm \dg )^{2} 
e^{-i \phi_{p}(t)}    \right),
    \label{eq:appHprime1}
\end{align}
where $\varphi_p(t) = 2 \int_0^t dt' \omega_p(t')$; here the spin frequency $\omega_{s}(t)$, the parametric drive frequency $\omega_{p}(t)$ and parametric drive amplitude $\lambda(t)$
are all time dependent.  This corresponds to chirping the parametric drive frequency, and also varying the spin Larmor frequency.
We point out, however, that instead of making the spin Larmor frequency be time-dependent, we could also arrive at the same effective Hamiltonian by instead introducing appropriate time-dependence to $\omega_{c}$.

Going into a rotating frame defined by 
$\Uop_{\rm rf}(t)=\exp \left[ i \varphi_p(t) (\cm \dg \cm +  \Sz)  \right]$, gives
\begin{align}
    \hat{H}_{\rm rf}(t) &=  \Uop_{\rm rf}(t) \hat{H}_{\rm lab}(t) 
    \Uop^{\dagger}_{\rm rf}(t)
    + i \dot{\hat{U}}_{\rm rf}(t) \Uop_{\rm rf}^{\dagger}(t) \nonumber \\
    &=    \Delta_c(t) \cm \dg \cm 
    + \Delta_s(t) \hat{S}_z 
    + \left( g \cm \dg \Sm + \frac{\lambda(t)}{2} \cm \dg \cm \dg + \hc \right),
    \label{eq:appHprimeRF}
\end{align}
with $\Delta_{c}(t) = \omega_{c} - \omega_{p}(t)$, 
$\Delta_{s}(t) = \omega_{s}(t) - \omega_{p}(t)$.

For our adiabatic protocol, we first pick a desired time dependent squeezing parameter $r(t)$ that evolves smoothly and slowly from $r(0)=0$ to $r(\tau_{\rm prot}) = r_f$, with $\tau_{\rm prot}$ representing the total protocol time. 
We also choose a target cavity Bogoliubov mode energy $E_{\beta}$ that is time-independent (as this helps minimize non-adiabatic errors \cite{DavisPechukas1976})
Once these choices are made, we use them to determine the time-dependence of the chirped parametric drive frequency $\omega_p(t)$ and the {\it real} part of the parametric drive amplitude $\lambda(t)$.  This is done via the equations:
\begin{align}
    \Delta_c(t) & = E_{\beta} \cosh 2r(t), \\
    \Re \lambda(t) & = E_{\beta} \sinh 2r(t).
\end{align}
Finally, we chose the imaginary part of the parametric drive amplitude that satisfies
\begin{align}
    \Im \lambda(t) & =  \dot{r}(t)
    \label{eq:lambdaToR}
\end{align}
With these choices, we can now further transform our system into a moving squeezed frame via the unitary
\begin{align}
    \Uop_{\rm sq}(t) &=  e^{-\frac{r(t)}{2} \left[  \cm^{2} - (\cm \dg)^{2} \right]}, 
    \label{eq:UprimeSq}
\end{align}
Our choices above ensure that the resulting instantaneous Hamiltonian in this moving frame has a cavity-only part that it diagonal; the choice of $\Im \lambda(t)$ cancels the inertial term associated with the transformation, which results in 
\begin{align}
    \hat{H}_{\rm sq}(t) &=  \Delta_s(t) \hat{S}_z  + E_{\beta} \hat{c}^\dagger \hat{c}  +  \left(g \hat{c} \dg \Sigmaop[r(t)] + \hc \right),
    \label{eq:appHprine3}
\end{align}
where we now make the $r$- (and therefore time-) dependence of the Bogoliubov spin operator $\Sigmaop[r(t)]$ explicit.

We now have our desired Hamiltonian: in this moving squeezed frame, the instantaneous Hamiltonian involves the cavity interacting with the instantaneous spin-Bogoliubov mode $\Sigmaop[r(t)]$.  We can now follow the same steps we took in the time-independent case, and eliminate the cavity via a Schreiffer-Wolff transformation to obtain the desired Eq.~(\ref{eq:Heff}) of the main text.  Note that the remaining freedom in choosing the time dependence of the spin frequency $\omega_s(t)$ (and hence $\Delta_s(t)$) can be used to cancel the final term $\sim \hat{S}_z$, as was done in the time-independent protocol.  

The attentive reader will of course note that the Schreiffer-Wolff transformation we need to make in the time-dependent case will involve a time-dependent generator, and will hence produce new terms compared to the static case.  However, these terms can be safely neglected as we are interested in an adiabatic protocol, where $r(t)$ evolves slowly.  In particular, we will have $\dot{r}(t) \ll \chi$, implying that 
terms associated with the time-derivative of the Schreiffer-Wolff generator do not contribute to the transformed Hamiltonian to leading order in $\chi$.  

\begin{figure}[t]
    \centering
    \includegraphics[width=0.5 \columnwidth]{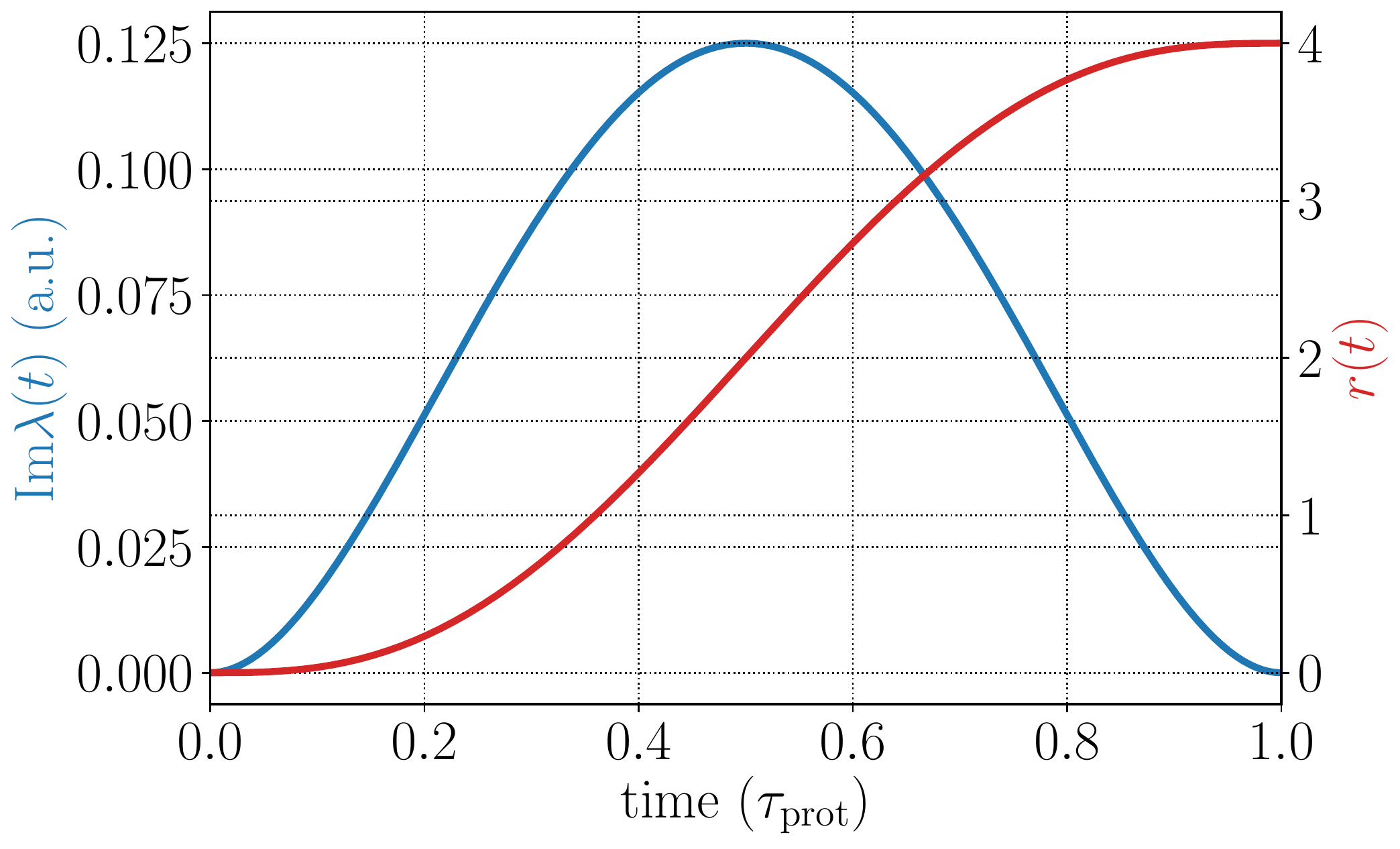}
    \vspace{-0.3cm}
    \caption{
        A plot showing $\Im \lambda(t)$ (blue, left axis) and the corresponding $r(t)$ (red, right axis) as a function of protocol time. This pulse was used in the simulations shown in Fig.~(4) (see main text for details). 
    \label{fig:adiabaticPulse}
    }
\end{figure}

{\it Adiabatic pulse--- } 
In order for our adiabatic protocol to work, we have to choose the drive $\lambda(t)$ so that $r(t)$ slowly increases from zero to its desired value $r_{f}$ at the end of the evolution. 
To do so, we choose a $\Im \lambda(t)$ that takes the form of a polynomial 
\begin{align}
    \Im \lambda(t)&=  \frac{30 r_{f}}{\tau_{\rm prot}} \left(\frac{t}{\tau_{\rm prot}}-1\right)^2\left(\frac{t}{\tau_{\rm prot}}\right)^2, 
    \label{eq:lambda(t)}
\end{align}
which is chosen to be smooth, with $\Im \lambda(0) = \Im \lambda(\tau_{\rm prot})= \dot \lambda(0) = \Im \dot\lambda(\tau_{\rm prot})=0$ \cite{wiebe2012improved}, and to satisfy $r(\tau_{\rm prot}) = r_{f}$.  
Using \cref{eq:lambdaToR}, lets us write $r(t)$ as
\begin{align}
    r(t) = \frac{r_{f}}{2} \left(\frac{3t}{\tau_{\rm prot}} \left(\frac{2t}{\tau_{\rm prot}}-5\right) + 10\right) \left(\frac{t}{\tau_{\rm prot}}\right)^3.
    \label{eq:rt}
\end{align}
Figure~(\ref{fig:adiabaticPulse}) shows $\Im \lambda(t)$ and $r(t)$ used in the simulations described in the main text, with $r_{f}=4$.

\section{Linearized theory}

In this section, we use the linearized theory to study the minimum squeezing parameter $\xi_R^2$ that can be achieved by the ITAT Hamiltonian in Eq.~(\ref{eq:appTAT1}).  We focus on the time evolution of the relevant spin operator variances, $\langle\Sy^2\rangle$, $\langle\Sz^2\rangle$, and $\langle\frac{\Sy\Sz+\Sz\Sy}{2}\rangle$, while assuming the other spin operator expectation values change negligibly during the evolution, i.e.,
\begin{equation}
    \langle\Sx\rangle=\frac{N}{2}~~,~~\langle\Sy\rangle=\langle\Sz\rangle=0~~,~~\langle\Sx^2\rangle=\frac{N^2}{4}~~,~~\langle\frac{\Sx\Sy+\Sy\Sx}{2}\rangle = \langle\frac{\Sx\Sz+\Sz\Sx}{2}\rangle=0~.
\end{equation}
These assumptions are consistent with starting the evolution with all the spins aligned along the positive $\Sx$ axis.
Furthermore, we can assume the dynamics of these variances are dominated by the Hamiltonian, while the dephasing and dissipation contribute only weakly. The equation of motion is then given by
\begin{equation}
    \frac{d}{dt}\left(\begin{array}{c}
        \langle\Sy^2\rangle \\
        \langle\Sz^2\rangle \\
        \langle\frac{\Sy\Sz+\Sz\Sy}{2}\rangle
    \end{array}\right) = 
    \frac{2}{3}\tilde{\chi}
    \left(\begin{array}{ccc}
        0 & 0 & 2N \\
        0 & 0 & 2N \\
        -N & -N & 0
    \end{array}\right)\left(\begin{array}{c}
        \langle\Sy^2\rangle \\
        \langle\Sz^2\rangle \\
        \langle\frac{\Sy\Sz+\Sz\Sy}{2}\rangle
    \end{array}\right) +
    \gamma_\phi\left(\begin{array}{c}
        \frac{N}{2} \\
        0 \\
        0
    \end{array}\right) +
    \Gamma\left(\begin{array}{c}
        0 \\
        e^{4r_0} \frac{N^2}{4} \\
        0
    \end{array}\right)~.
\end{equation}
The time evolution of the variances can be solved analytically. We find that the solution involves one exponentially increasing eigenvector and one exponentially decreasing eigenvector.  It means that under the ITAT interaction, one spin direction is anti-squeezed and another direction is squeezed.

The spin variance along any direction perpendicular to the mean spin direction of the collective spin, i.e. $\hat{S}_\theta\equiv \cos\theta \Sy + \sin\theta\Sz$, is given by
\begin{equation}
    \langle \hat{S}_{\theta}^{2}\rangle = \cos^2\theta\langle \Sy^2\rangle + \sin^2\theta \langle\Sz^2\rangle + 2 \sin\theta \cos\theta \langle \frac{\Sy\Sz+\Sz\Sy}{2}\rangle~.
\end{equation}
In the case we are studying, the direction with the minimum spin variance corresponds to the direction where the variance is squeezed. We find that this direction is $\theta=-\pi/4$, and the dynamics of this variance follows
\begin{equation}
    \frac{d}{dt}\langle \hat{S}^2_{-\pi/4}\rangle = -\frac{4}{3}N \tilde{\chi}\langle\hat{S}^2_{-\pi/4}\rangle + \frac{N}{4}\gamma_\phi +  \frac{N^2}{4}\Gamma~.
\end{equation}
This spin variance reduces monotonically during evolution, and therefore the minimum variance is attained at the steady state:
\begin{equation}
    \langle \Delta \hat{S}^2_\perp\rangle= \langle\hat{S}^2_{-\pi/4}(t\rightarrow \infty)\rangle = \frac{3}{16}\frac{\gamma_\phi + N \Gamma}{\tilde{\chi}}~.
\end{equation}

Here both the ITAT interaction strength $\tilde{\chi}$ and cavity-induced spin dissipation rate $\Gamma$ depend on the Bogoliubov mode energy $E_\beta$:
\begin{equation}
    \tilde{\chi}=\frac{3}{2\sqrt{2}}\frac{g^2}{E_\beta},~~\Gamma = \kappa \frac{g^2}{E_\beta^2}~.
\end{equation}
Finally, $\langle \Delta\hat{S}_\perp^2\rangle$ can be minimized by choosing the optimal Bogoliubov mode energy 
\begin{equation}
    E_\beta = \sqrt{N}\sqrt{\frac{g^2 \kappa}{\gamma_\phi}}~.
\end{equation}
The minimum squeezing parameter is given by
\begin{equation}
    \textrm{min}\{\xi_R^2\} = \sqrt{\frac{2}{\mathcal{C}}}~,
\end{equation}
where $\mathcal{C}\equiv N g^2/(\kappa \gamma_\phi)$ is the collective spin cooperativity.

\bibliography{library}

\begin{thebibliography}{33}%
\makeatletter
\providecommand \@ifxundefined [1]{%
 \@ifx{#1\undefined}
}%
\providecommand \@ifnum [1]{%
 \ifnum #1\expandafter \@firstoftwo
 \else \expandafter \@secondoftwo
 \fi
}%
\providecommand \@ifx [1]{%
 \ifx #1\expandafter \@firstoftwo
 \else \expandafter \@secondoftwo
 \fi
}%
\providecommand \natexlab [1]{#1}%
\providecommand \enquote  [1]{``#1''}%
\providecommand \bibnamefont  [1]{#1}%
\providecommand \bibfnamefont [1]{#1}%
\providecommand \citenamefont [1]{#1}%
\providecommand \href@noop [0]{\@secondoftwo}%
\providecommand \href [0]{\begingroup \@sanitize@url \@href}%
\providecommand \@href[1]{\@@startlink{#1}\@@href}%
\providecommand \@@href[1]{\endgroup#1\@@endlink}%
\providecommand \@sanitize@url [0]{\catcode `\\12\catcode `\$12\catcode
  `\&12\catcode `\#12\catcode `\^12\catcode `\_12\catcode `\%12\relax}%
\providecommand \@@startlink[1]{}%
\providecommand \@@endlink[0]{}%
\providecommand \url  [0]{\begingroup\@sanitize@url \@url }%
\providecommand \@url [1]{\endgroup\@href {#1}{\urlprefix }}%
\providecommand \urlprefix  [0]{URL }%
\providecommand \Eprint [0]{\href }%
\providecommand \doibase [0]{http://dx.doi.org/}%
\providecommand \selectlanguage [0]{\@gobble}%
\providecommand \bibinfo  [0]{\@secondoftwo}%
\providecommand \bibfield  [0]{\@secondoftwo}%
\providecommand \translation [1]{[#1]}%
\providecommand \BibitemOpen [0]{}%
\providecommand \bibitemStop [0]{}%
\providecommand \bibitemNoStop [0]{.\EOS\space}%
\providecommand \EOS [0]{\spacefactor3000\relax}%
\providecommand \BibitemShut  [1]{\csname bibitem#1\endcsname}%
\let\auto@bib@innerbib\@empty
\bibitem [{\citenamefont {Ma}\ \emph {et~al.}(2011)\citenamefont {Ma},
  \citenamefont {Wang}, \citenamefont {Sun},\ and\ \citenamefont
  {Nori}}]{ma2011quantum}%
  \BibitemOpen
  \bibfield  {author} {\bibinfo {author} {\bibfnamefont {J.}~\bibnamefont
  {Ma}}, \bibinfo {author} {\bibfnamefont {X.}~\bibnamefont {Wang}}, \bibinfo
  {author} {\bibfnamefont {C.-P.}\ \bibnamefont {Sun}}, \ and\ \bibinfo
  {author} {\bibfnamefont {F.}~\bibnamefont {Nori}},\ }\href@noop {} {\bibfield
   {journal} {\bibinfo  {journal} {Physics Reports}\ }\textbf {\bibinfo
  {volume} {509}},\ \bibinfo {pages} {89} (\bibinfo {year} {2011})}\BibitemShut
  {NoStop}%
\bibitem [{\citenamefont {Kitagawa}\ and\ \citenamefont
  {Ueda}(1993)}]{kitagawa1993squeezed}%
  \BibitemOpen
  \bibfield  {author} {\bibinfo {author} {\bibfnamefont {M.}~\bibnamefont
  {Kitagawa}}\ and\ \bibinfo {author} {\bibfnamefont {M.}~\bibnamefont
  {Ueda}},\ }\href {\doibase 10.1103/PhysRevA.47.5138} {\bibfield  {journal}
  {\bibinfo  {journal} {Phys. Rev. A}\ }\textbf {\bibinfo {volume} {47}},\
  \bibinfo {pages} {5138} (\bibinfo {year} {1993})}\BibitemShut {NoStop}%
\bibitem [{\citenamefont {Agarwal}\ and\ \citenamefont
  {Puri}(1994)}]{AgarwalPuri1994}%
  \BibitemOpen
  \bibfield  {author} {\bibinfo {author} {\bibfnamefont {G.~S.}\ \bibnamefont
  {Agarwal}}\ and\ \bibinfo {author} {\bibfnamefont {R.~R.}\ \bibnamefont
  {Puri}},\ }\href {\doibase 10.1103/PhysRevA.49.4968} {\bibfield  {journal}
  {\bibinfo  {journal} {Phys. Rev. A}\ }\textbf {\bibinfo {volume} {49}},\
  \bibinfo {pages} {4968} (\bibinfo {year} {1994})}\BibitemShut {NoStop}%
\bibitem [{\citenamefont {Schleier-Smith}\ \emph {et~al.}(2010)\citenamefont
  {Schleier-Smith}, \citenamefont {Leroux},\ and\ \citenamefont
  {Vuleti\ifmmode~\acute{c}\else \'{c}\fi{}}}]{schleier2010squeezing}%
  \BibitemOpen
  \bibfield  {author} {\bibinfo {author} {\bibfnamefont {M.~H.}\ \bibnamefont
  {Schleier-Smith}}, \bibinfo {author} {\bibfnamefont {I.~D.}\ \bibnamefont
  {Leroux}}, \ and\ \bibinfo {author} {\bibfnamefont {V.}~\bibnamefont
  {Vuleti\ifmmode~\acute{c}\else \'{c}\fi{}}},\ }\href {\doibase
  10.1103/PhysRevA.81.021804} {\bibfield  {journal} {\bibinfo  {journal} {Phys.
  Rev. A}\ }\textbf {\bibinfo {volume} {81}},\ \bibinfo {pages} {021804}
  (\bibinfo {year} {2010})}\BibitemShut {NoStop}%
\bibitem [{\citenamefont {Bennett}\ \emph {et~al.}(2013)\citenamefont
  {Bennett}, \citenamefont {Yao}, \citenamefont {Otterbach}, \citenamefont
  {Zoller}, \citenamefont {Rabl},\ and\ \citenamefont
  {Lukin}}]{bennett2013phonon}%
  \BibitemOpen
  \bibfield  {author} {\bibinfo {author} {\bibfnamefont {S.~D.}\ \bibnamefont
  {Bennett}}, \bibinfo {author} {\bibfnamefont {N.~Y.}\ \bibnamefont {Yao}},
  \bibinfo {author} {\bibfnamefont {J.}~\bibnamefont {Otterbach}}, \bibinfo
  {author} {\bibfnamefont {P.}~\bibnamefont {Zoller}}, \bibinfo {author}
  {\bibfnamefont {P.}~\bibnamefont {Rabl}}, \ and\ \bibinfo {author}
  {\bibfnamefont {M.~D.}\ \bibnamefont {Lukin}},\ }\href {\doibase
  10.1103/PhysRevLett.110.156402} {\bibfield  {journal} {\bibinfo  {journal}
  {Phys. Rev. Lett.}\ }\textbf {\bibinfo {volume} {110}},\ \bibinfo {pages}
  {156402} (\bibinfo {year} {2013})}\BibitemShut {NoStop}%
\bibitem [{\citenamefont {Dalla~Torre}\ \emph {et~al.}(2013)\citenamefont
  {Dalla~Torre}, \citenamefont {Otterbach}, \citenamefont {Demler},
  \citenamefont {Vuletic},\ and\ \citenamefont {Lukin}}]{dalla2013dissipative}%
  \BibitemOpen
  \bibfield  {author} {\bibinfo {author} {\bibfnamefont {E.~G.}\ \bibnamefont
  {Dalla~Torre}}, \bibinfo {author} {\bibfnamefont {J.}~\bibnamefont
  {Otterbach}}, \bibinfo {author} {\bibfnamefont {E.}~\bibnamefont {Demler}},
  \bibinfo {author} {\bibfnamefont {V.}~\bibnamefont {Vuletic}}, \ and\
  \bibinfo {author} {\bibfnamefont {M.~D.}\ \bibnamefont {Lukin}},\ }\href
  {\doibase 10.1103/PhysRevLett.110.120402} {\bibfield  {journal} {\bibinfo
  {journal} {Phys. Rev. Lett.}\ }\textbf {\bibinfo {volume} {110}},\ \bibinfo
  {pages} {120402} (\bibinfo {year} {2013})}\BibitemShut {NoStop}%
\bibitem [{\citenamefont {Hu}\ \emph {et~al.}(2017)\citenamefont {Hu},
  \citenamefont {Chen}, \citenamefont {Vendeiro}, \citenamefont {Urvoy},
  \citenamefont {Braverman},\ and\ \citenamefont {Vuleti\ifmmode~\acute{c}\else
  \'{c}\fi{}}}]{hu2017vacuum}%
  \BibitemOpen
  \bibfield  {author} {\bibinfo {author} {\bibfnamefont {J.}~\bibnamefont
  {Hu}}, \bibinfo {author} {\bibfnamefont {W.}~\bibnamefont {Chen}}, \bibinfo
  {author} {\bibfnamefont {Z.}~\bibnamefont {Vendeiro}}, \bibinfo {author}
  {\bibfnamefont {A.}~\bibnamefont {Urvoy}}, \bibinfo {author} {\bibfnamefont
  {B.}~\bibnamefont {Braverman}}, \ and\ \bibinfo {author} {\bibfnamefont
  {V.}~\bibnamefont {Vuleti\ifmmode~\acute{c}\else \'{c}\fi{}}},\ }\href
  {\doibase 10.1103/PhysRevA.96.050301} {\bibfield  {journal} {\bibinfo
  {journal} {Phys. Rev. A}\ }\textbf {\bibinfo {volume} {96}},\ \bibinfo
  {pages} {050301} (\bibinfo {year} {2017})}\BibitemShut {NoStop}%
\bibitem [{\citenamefont {Lewis-Swan}\ \emph {et~al.}(2018)\citenamefont
  {Lewis-Swan}, \citenamefont {Norcia}, \citenamefont {Cline}, \citenamefont
  {Thompson},\ and\ \citenamefont {Rey}}]{lewis2018robust}%
  \BibitemOpen
  \bibfield  {author} {\bibinfo {author} {\bibfnamefont {R.~J.}\ \bibnamefont
  {Lewis-Swan}}, \bibinfo {author} {\bibfnamefont {M.~A.}\ \bibnamefont
  {Norcia}}, \bibinfo {author} {\bibfnamefont {J.~R.~K.}\ \bibnamefont
  {Cline}}, \bibinfo {author} {\bibfnamefont {J.~K.}\ \bibnamefont {Thompson}},
  \ and\ \bibinfo {author} {\bibfnamefont {A.~M.}\ \bibnamefont {Rey}},\ }\href
  {\doibase 10.1103/PhysRevLett.121.070403} {\bibfield  {journal} {\bibinfo
  {journal} {Phys. Rev. Lett.}\ }\textbf {\bibinfo {volume} {121}},\ \bibinfo
  {pages} {070403} (\bibinfo {year} {2018})}\BibitemShut {NoStop}%
\bibitem [{\citenamefont {Leroux}\ \emph {et~al.}(2010)\citenamefont {Leroux},
  \citenamefont {Schleier-Smith},\ and\ \citenamefont
  {Vuleti\ifmmode~\acute{c}\else \'{c}\fi{}}}]{LerouxPRL2010}%
  \BibitemOpen
  \bibfield  {author} {\bibinfo {author} {\bibfnamefont {I.~D.}\ \bibnamefont
  {Leroux}}, \bibinfo {author} {\bibfnamefont {M.~H.}\ \bibnamefont
  {Schleier-Smith}}, \ and\ \bibinfo {author} {\bibfnamefont {V.}~\bibnamefont
  {Vuleti\ifmmode~\acute{c}\else \'{c}\fi{}}},\ }\href {\doibase
  10.1103/PhysRevLett.104.073602} {\bibfield  {journal} {\bibinfo  {journal}
  {Phys. Rev. Lett.}\ }\textbf {\bibinfo {volume} {104}},\ \bibinfo {pages}
  {073602} (\bibinfo {year} {2010})}\BibitemShut {NoStop}%
\bibitem [{\citenamefont {Riedel}\ \emph {et~al.}(2010)\citenamefont {Riedel},
  \citenamefont {B{\"o}hi}, \citenamefont {Li}, \citenamefont {H{\"a}nsch},
  \citenamefont {Sinatra},\ and\ \citenamefont {Treutlein}}]{Riedel2010}%
  \BibitemOpen
  \bibfield  {author} {\bibinfo {author} {\bibfnamefont {M.~F.}\ \bibnamefont
  {Riedel}}, \bibinfo {author} {\bibfnamefont {P.}~\bibnamefont {B{\"o}hi}},
  \bibinfo {author} {\bibfnamefont {Y.}~\bibnamefont {Li}}, \bibinfo {author}
  {\bibfnamefont {T.~W.}\ \bibnamefont {H{\"a}nsch}}, \bibinfo {author}
  {\bibfnamefont {A.}~\bibnamefont {Sinatra}}, \ and\ \bibinfo {author}
  {\bibfnamefont {P.}~\bibnamefont {Treutlein}},\ }\href@noop {} {\bibfield
  {journal} {\bibinfo  {journal} {Nature}\ }\textbf {\bibinfo {volume} {464}},\
  \bibinfo {pages} {1170} (\bibinfo {year} {2010})}\BibitemShut {NoStop}%
\bibitem [{\citenamefont {Gross}\ \emph {et~al.}(2010)\citenamefont {Gross},
  \citenamefont {Zibold}, \citenamefont {Nicklas}, \citenamefont {Est{\`e}ve},\
  and\ \citenamefont {Oberthaler}}]{GrossNature2010}%
  \BibitemOpen
  \bibfield  {author} {\bibinfo {author} {\bibfnamefont {C.}~\bibnamefont
  {Gross}}, \bibinfo {author} {\bibfnamefont {T.}~\bibnamefont {Zibold}},
  \bibinfo {author} {\bibfnamefont {E.}~\bibnamefont {Nicklas}}, \bibinfo
  {author} {\bibfnamefont {J.}~\bibnamefont {Est{\`e}ve}}, \ and\ \bibinfo
  {author} {\bibfnamefont {M.~K.}\ \bibnamefont {Oberthaler}},\ }\href
  {\doibase 10.1038/nature08919} {\bibfield  {journal} {\bibinfo  {journal}
  {Nature}\ }\textbf {\bibinfo {volume} {464}},\ \bibinfo {pages} {1165}
  (\bibinfo {year} {2010})}\BibitemShut {NoStop}%
\bibitem [{\citenamefont {Hosten}\ \emph {et~al.}(2016)\citenamefont {Hosten},
  \citenamefont {Krishnakumar}, \citenamefont {Engelsen},\ and\ \citenamefont
  {Kasevich}}]{HostenScience2016}%
  \BibitemOpen
  \bibfield  {author} {\bibinfo {author} {\bibfnamefont {O.}~\bibnamefont
  {Hosten}}, \bibinfo {author} {\bibfnamefont {R.}~\bibnamefont
  {Krishnakumar}}, \bibinfo {author} {\bibfnamefont {N.~J.}\ \bibnamefont
  {Engelsen}}, \ and\ \bibinfo {author} {\bibfnamefont {M.~A.}\ \bibnamefont
  {Kasevich}},\ }\href {\doibase 10.1126/science.aaf3397} {\bibfield  {journal}
  {\bibinfo  {journal} {Science}\ }\textbf {\bibinfo {volume} {352}},\ \bibinfo
  {pages} {1552} (\bibinfo {year} {2016})}\BibitemShut {NoStop}%
\bibitem [{\citenamefont {Andr\'e}\ and\ \citenamefont
  {Lukin}(2002)}]{andre2002atom}%
  \BibitemOpen
  \bibfield  {author} {\bibinfo {author} {\bibfnamefont {A.}~\bibnamefont
  {Andr\'e}}\ and\ \bibinfo {author} {\bibfnamefont {M.~D.}\ \bibnamefont
  {Lukin}},\ }\href {\doibase 10.1103/PhysRevA.65.053819} {\bibfield  {journal}
  {\bibinfo  {journal} {Phys. Rev. A}\ }\textbf {\bibinfo {volume} {65}},\
  \bibinfo {pages} {053819} (\bibinfo {year} {2002})}\BibitemShut {NoStop}%
\bibitem [{\citenamefont {Pezz\`e}\ \emph {et~al.}(2018)\citenamefont
  {Pezz\`e}, \citenamefont {Smerzi}, \citenamefont {Oberthaler}, \citenamefont
  {Schmied},\ and\ \citenamefont {Treutlein}}]{pezze2018quantum}%
  \BibitemOpen
  \bibfield  {author} {\bibinfo {author} {\bibfnamefont {L.}~\bibnamefont
  {Pezz\`e}}, \bibinfo {author} {\bibfnamefont {A.}~\bibnamefont {Smerzi}},
  \bibinfo {author} {\bibfnamefont {M.~K.}\ \bibnamefont {Oberthaler}},
  \bibinfo {author} {\bibfnamefont {R.}~\bibnamefont {Schmied}}, \ and\
  \bibinfo {author} {\bibfnamefont {P.}~\bibnamefont {Treutlein}},\ }\href
  {\doibase 10.1103/RevModPhys.90.035005} {\bibfield  {journal} {\bibinfo
  {journal} {Rev. Mod. Phys.}\ }\textbf {\bibinfo {volume} {90}},\ \bibinfo
  {pages} {035005} (\bibinfo {year} {2018})}\BibitemShut {NoStop}%
\bibitem [{\citenamefont {Lee}\ \emph {et~al.}(2017)\citenamefont {Lee},
  \citenamefont {Lee}, \citenamefont {Cady}, \citenamefont {Ovartchaiyapong},\
  and\ \citenamefont {Jayich}}]{JayichReview2017}%
  \BibitemOpen
  \bibfield  {author} {\bibinfo {author} {\bibfnamefont {D.}~\bibnamefont
  {Lee}}, \bibinfo {author} {\bibfnamefont {K.~W.}\ \bibnamefont {Lee}},
  \bibinfo {author} {\bibfnamefont {J.~V.}\ \bibnamefont {Cady}}, \bibinfo
  {author} {\bibfnamefont {P.}~\bibnamefont {Ovartchaiyapong}}, \ and\ \bibinfo
  {author} {\bibfnamefont {A.~C.~B.}\ \bibnamefont {Jayich}},\ }\href@noop {}
  {\bibfield  {journal} {\bibinfo  {journal} {Journal of Optics}\ }\textbf
  {\bibinfo {volume} {19}},\ \bibinfo {pages} {033001} (\bibinfo {year}
  {2017})}\BibitemShut {NoStop}%
\bibitem [{\citenamefont {Bienfait}\ \emph {et~al.}(2017)\citenamefont
  {Bienfait}, \citenamefont {Campagne-Ibarcq}, \citenamefont {Kiilerich},
  \citenamefont {Zhou}, \citenamefont {Probst}, \citenamefont {Pla},
  \citenamefont {Schenkel}, \citenamefont {Vion}, \citenamefont {Esteve},
  \citenamefont {Morton}, \citenamefont {Moelmer},\ and\ \citenamefont
  {Bertet}}]{BertetPRX2017}%
  \BibitemOpen
  \bibfield  {author} {\bibinfo {author} {\bibfnamefont {A.}~\bibnamefont
  {Bienfait}}, \bibinfo {author} {\bibfnamefont {P.}~\bibnamefont
  {Campagne-Ibarcq}}, \bibinfo {author} {\bibfnamefont {A.~H.}\ \bibnamefont
  {Kiilerich}}, \bibinfo {author} {\bibfnamefont {X.}~\bibnamefont {Zhou}},
  \bibinfo {author} {\bibfnamefont {S.}~\bibnamefont {Probst}}, \bibinfo
  {author} {\bibfnamefont {J.~J.}\ \bibnamefont {Pla}}, \bibinfo {author}
  {\bibfnamefont {T.}~\bibnamefont {Schenkel}}, \bibinfo {author}
  {\bibfnamefont {D.}~\bibnamefont {Vion}}, \bibinfo {author} {\bibfnamefont
  {D.}~\bibnamefont {Esteve}}, \bibinfo {author} {\bibfnamefont {J.~J.~L.}\
  \bibnamefont {Morton}}, \bibinfo {author} {\bibfnamefont {K.}~\bibnamefont
  {Moelmer}}, \ and\ \bibinfo {author} {\bibfnamefont {P.}~\bibnamefont
  {Bertet}},\ }\href {\doibase 10.1103/PhysRevX.7.041011} {\bibfield  {journal}
  {\bibinfo  {journal} {Phys. Rev. X}\ }\textbf {\bibinfo {volume} {7}},\
  \bibinfo {pages} {041011} (\bibinfo {year} {2017})}\BibitemShut {NoStop}%
\bibitem [{\citenamefont {Ge}\ \emph {et~al.}(2019)\citenamefont {Ge},
  \citenamefont {Sawyer}, \citenamefont {Britton}, \citenamefont {Jacobs},
  \citenamefont {Bollinger},\ and\ \citenamefont
  {Foss-Feig}}]{FossFeigPRL2019}%
  \BibitemOpen
  \bibfield  {author} {\bibinfo {author} {\bibfnamefont {W.}~\bibnamefont
  {Ge}}, \bibinfo {author} {\bibfnamefont {B.~C.}\ \bibnamefont {Sawyer}},
  \bibinfo {author} {\bibfnamefont {J.~W.}\ \bibnamefont {Britton}}, \bibinfo
  {author} {\bibfnamefont {K.}~\bibnamefont {Jacobs}}, \bibinfo {author}
  {\bibfnamefont {J.~J.}\ \bibnamefont {Bollinger}}, \ and\ \bibinfo {author}
  {\bibfnamefont {M.}~\bibnamefont {Foss-Feig}},\ }\href {\doibase
  10.1103/PhysRevLett.122.030501} {\bibfield  {journal} {\bibinfo  {journal}
  {Phys. Rev. Lett.}\ }\textbf {\bibinfo {volume} {122}},\ \bibinfo {pages}
  {030501} (\bibinfo {year} {2019})}\BibitemShut {NoStop}%
\bibitem [{\citenamefont {Qin}\ \emph {et~al.}(2019)\citenamefont {Qin},
  \citenamefont {Chen}, \citenamefont {Wang}, \citenamefont {Miranowicz},\ and\
  \citenamefont {Nori}}]{NoriParametric2019}%
  \BibitemOpen
  \bibfield  {author} {\bibinfo {author} {\bibfnamefont {W.}~\bibnamefont
  {Qin}}, \bibinfo {author} {\bibfnamefont {Y.-H.}\ \bibnamefont {Chen}},
  \bibinfo {author} {\bibfnamefont {X.}~\bibnamefont {Wang}}, \bibinfo {author}
  {\bibfnamefont {A.}~\bibnamefont {Miranowicz}}, \ and\ \bibinfo {author}
  {\bibfnamefont {F.}~\bibnamefont {Nori}},\ }\href@noop {} {\bibfield
  {journal} {\bibinfo  {journal} {arXiv:1912.04039v1}\ } (\bibinfo {year}
  {2019})}\BibitemShut {NoStop}%
\bibitem [{\citenamefont {Agarwal}\ and\ \citenamefont
  {Puri}(1990)}]{AgarwalPuri1990}%
  \BibitemOpen
  \bibfield  {author} {\bibinfo {author} {\bibfnamefont {G.~S.}\ \bibnamefont
  {Agarwal}}\ and\ \bibinfo {author} {\bibfnamefont {R.~R.}\ \bibnamefont
  {Puri}},\ }\href {\doibase 10.1103/PhysRevA.41.3782} {\bibfield  {journal}
  {\bibinfo  {journal} {Phys. Rev. A}\ }\textbf {\bibinfo {volume} {41}},\
  \bibinfo {pages} {3782} (\bibinfo {year} {1990})}\BibitemShut {NoStop}%
\bibitem [{sup()}]{supp}%
  \BibitemOpen
  \href@noop {} {}\bibinfo {note} {See supplementary information}\BibitemShut
  {NoStop}%
\bibitem [{\citenamefont {Agarwal}\ \emph {et~al.}(1997)\citenamefont
  {Agarwal}, \citenamefont {Puri},\ and\ \citenamefont
  {Singh}}]{AgarwalPRA1997}%
  \BibitemOpen
  \bibfield  {author} {\bibinfo {author} {\bibfnamefont {G.~S.}\ \bibnamefont
  {Agarwal}}, \bibinfo {author} {\bibfnamefont {R.~R.}\ \bibnamefont {Puri}}, \
  and\ \bibinfo {author} {\bibfnamefont {R.~P.}\ \bibnamefont {Singh}},\ }\href
  {\doibase 10.1103/PhysRevA.56.2249} {\bibfield  {journal} {\bibinfo
  {journal} {Phys. Rev. A}\ }\textbf {\bibinfo {volume} {56}},\ \bibinfo
  {pages} {2249} (\bibinfo {year} {1997})}\BibitemShut {NoStop}%
\bibitem [{\citenamefont {Cappellaro}\ and\ \citenamefont
  {Lukin}(2009)}]{Cappellaro2009}%
  \BibitemOpen
  \bibfield  {author} {\bibinfo {author} {\bibfnamefont {P.}~\bibnamefont
  {Cappellaro}}\ and\ \bibinfo {author} {\bibfnamefont {M.~D.}\ \bibnamefont
  {Lukin}},\ }\href {\doibase 10.1103/PhysRevA.80.032311} {\bibfield  {journal}
  {\bibinfo  {journal} {Phys. Rev. A}\ }\textbf {\bibinfo {volume} {80}},\
  \bibinfo {pages} {032311} (\bibinfo {year} {2009})}\BibitemShut {NoStop}%
\bibitem [{\citenamefont {Liu}\ \emph {et~al.}(2011)\citenamefont {Liu},
  \citenamefont {Xu}, \citenamefont {Jin},\ and\ \citenamefont
  {You}}]{You2011TAT}%
  \BibitemOpen
  \bibfield  {author} {\bibinfo {author} {\bibfnamefont {Y.~C.}\ \bibnamefont
  {Liu}}, \bibinfo {author} {\bibfnamefont {Z.~F.}\ \bibnamefont {Xu}},
  \bibinfo {author} {\bibfnamefont {G.~R.}\ \bibnamefont {Jin}}, \ and\
  \bibinfo {author} {\bibfnamefont {L.}~\bibnamefont {You}},\ }\href {\doibase
  10.1103/PhysRevLett.107.013601} {\bibfield  {journal} {\bibinfo  {journal}
  {Phys. Rev. Lett.}\ }\textbf {\bibinfo {volume} {107}},\ \bibinfo {pages}
  {013601} (\bibinfo {year} {2011})}\BibitemShut {NoStop}%
\bibitem [{\citenamefont {Borregaard}\ \emph {et~al.}(2017)\citenamefont
  {Borregaard}, \citenamefont {Davis}, \citenamefont {Bentsen}, \citenamefont
  {Schleier-Smith},\ and\ \citenamefont {S{\o}rensen}}]{borregaard2017one}%
  \BibitemOpen
  \bibfield  {author} {\bibinfo {author} {\bibfnamefont {J.}~\bibnamefont
  {Borregaard}}, \bibinfo {author} {\bibfnamefont {E.}~\bibnamefont {Davis}},
  \bibinfo {author} {\bibfnamefont {G.~S.}\ \bibnamefont {Bentsen}}, \bibinfo
  {author} {\bibfnamefont {M.~H.}\ \bibnamefont {Schleier-Smith}}, \ and\
  \bibinfo {author} {\bibfnamefont {A.~S.}\ \bibnamefont {S{\o}rensen}},\
  }\href@noop {} {\bibfield  {journal} {\bibinfo  {journal} {New J. Phys.}\
  }\textbf {\bibinfo {volume} {19}},\ \bibinfo {pages} {093021} (\bibinfo
  {year} {2017})}\BibitemShut {NoStop}%
\bibitem [{\citenamefont {Macri}\ \emph {et~al.}(2019)\citenamefont {Macri},
  \citenamefont {Nori}, \citenamefont {Savasta},\ and\ \citenamefont
  {Zueco}}]{NoriTATPreprint}%
  \BibitemOpen
  \bibfield  {author} {\bibinfo {author} {\bibfnamefont {V.}~\bibnamefont
  {Macri}}, \bibinfo {author} {\bibfnamefont {F.}~\bibnamefont {Nori}},
  \bibinfo {author} {\bibfnamefont {S.}~\bibnamefont {Savasta}}, \ and\
  \bibinfo {author} {\bibfnamefont {D.}~\bibnamefont {Zueco}},\ }\href@noop {}
  {\bibfield  {journal} {\bibinfo  {journal} {arXiv:1902.10377v1}\ } (\bibinfo
  {year} {2019})}\BibitemShut {NoStop}%
\bibitem [{\citenamefont {Strobel}\ \emph {et~al.}(2014)\citenamefont
  {Strobel}, \citenamefont {Muessel}, \citenamefont {Linnemann}, \citenamefont
  {Zibold}, \citenamefont {Hume}, \citenamefont {Pezz{\`e}}, \citenamefont
  {Smerzi},\ and\ \citenamefont {Oberthaler}}]{Strobel2014}%
  \BibitemOpen
  \bibfield  {author} {\bibinfo {author} {\bibfnamefont {H.}~\bibnamefont
  {Strobel}}, \bibinfo {author} {\bibfnamefont {W.}~\bibnamefont {Muessel}},
  \bibinfo {author} {\bibfnamefont {D.}~\bibnamefont {Linnemann}}, \bibinfo
  {author} {\bibfnamefont {T.}~\bibnamefont {Zibold}}, \bibinfo {author}
  {\bibfnamefont {D.~B.}\ \bibnamefont {Hume}}, \bibinfo {author}
  {\bibfnamefont {L.}~\bibnamefont {Pezz{\`e}}}, \bibinfo {author}
  {\bibfnamefont {A.}~\bibnamefont {Smerzi}}, \ and\ \bibinfo {author}
  {\bibfnamefont {M.~K.}\ \bibnamefont {Oberthaler}},\ }\href@noop {}
  {\bibfield  {journal} {\bibinfo  {journal} {Science}\ }\textbf {\bibinfo
  {volume} {345}},\ \bibinfo {pages} {424} (\bibinfo {year}
  {2014})}\BibitemShut {NoStop}%
\bibitem [{\citenamefont {Muessel}\ \emph {et~al.}(2015)\citenamefont
  {Muessel}, \citenamefont {Strobel}, \citenamefont {Linnemann}, \citenamefont
  {Zibold}, \citenamefont {Juli\'a-D\'{\i}az},\ and\ \citenamefont
  {Oberthaler}}]{Oberthaler2015}%
  \BibitemOpen
  \bibfield  {author} {\bibinfo {author} {\bibfnamefont {W.}~\bibnamefont
  {Muessel}}, \bibinfo {author} {\bibfnamefont {H.}~\bibnamefont {Strobel}},
  \bibinfo {author} {\bibfnamefont {D.}~\bibnamefont {Linnemann}}, \bibinfo
  {author} {\bibfnamefont {T.}~\bibnamefont {Zibold}}, \bibinfo {author}
  {\bibfnamefont {B.}~\bibnamefont {Juli\'a-D\'{\i}az}}, \ and\ \bibinfo
  {author} {\bibfnamefont {M.~K.}\ \bibnamefont {Oberthaler}},\ }\href
  {\doibase 10.1103/PhysRevA.92.023603} {\bibfield  {journal} {\bibinfo
  {journal} {Phys. Rev. A}\ }\textbf {\bibinfo {volume} {92}},\ \bibinfo
  {pages} {023603} (\bibinfo {year} {2015})}\BibitemShut {NoStop}%
\bibitem [{\citenamefont {Wineland}\ \emph {et~al.}(1994)\citenamefont
  {Wineland}, \citenamefont {Bollinger}, \citenamefont {Itano},\ and\
  \citenamefont {Heinzen}}]{wineland1994squeezed}%
  \BibitemOpen
  \bibfield  {author} {\bibinfo {author} {\bibfnamefont {D.~J.}\ \bibnamefont
  {Wineland}}, \bibinfo {author} {\bibfnamefont {J.~J.}\ \bibnamefont
  {Bollinger}}, \bibinfo {author} {\bibfnamefont {W.~M.}\ \bibnamefont
  {Itano}}, \ and\ \bibinfo {author} {\bibfnamefont {D.~J.}\ \bibnamefont
  {Heinzen}},\ }\href {\doibase 10.1103/PhysRevA.50.67} {\bibfield  {journal}
  {\bibinfo  {journal} {Phys. Rev. A}\ }\textbf {\bibinfo {volume} {50}},\
  \bibinfo {pages} {67} (\bibinfo {year} {1994})}\BibitemShut {NoStop}%
\bibitem [{\citenamefont {Johansson}\ \emph {et~al.}(2013)\citenamefont
  {Johansson}, \citenamefont {Nation},\ and\ \citenamefont
  {Nori}}]{johansson2013qutip}%
  \BibitemOpen
  \bibfield  {author} {\bibinfo {author} {\bibfnamefont {J.}~\bibnamefont
  {Johansson}}, \bibinfo {author} {\bibfnamefont {P.}~\bibnamefont {Nation}}, \
  and\ \bibinfo {author} {\bibfnamefont {F.}~\bibnamefont {Nori}},\ }\href@noop
  {} {\bibfield  {journal} {\bibinfo  {journal} {Computer Physics
  Communications}\ }\textbf {\bibinfo {volume} {184}},\ \bibinfo {pages} {1234}
  (\bibinfo {year} {2013})}\BibitemShut {NoStop}%
\bibitem [{\citenamefont {Shammah}\ \emph {et~al.}(2018)\citenamefont
  {Shammah}, \citenamefont {Ahmed}, \citenamefont {Lambert}, \citenamefont
  {De~Liberato},\ and\ \citenamefont {Nori}}]{shammah2018open}%
  \BibitemOpen
  \bibfield  {author} {\bibinfo {author} {\bibfnamefont {N.}~\bibnamefont
  {Shammah}}, \bibinfo {author} {\bibfnamefont {S.}~\bibnamefont {Ahmed}},
  \bibinfo {author} {\bibfnamefont {N.}~\bibnamefont {Lambert}}, \bibinfo
  {author} {\bibfnamefont {S.}~\bibnamefont {De~Liberato}}, \ and\ \bibinfo
  {author} {\bibfnamefont {F.}~\bibnamefont {Nori}},\ }\href@noop {} {\bibfield
   {journal} {\bibinfo  {journal} {Physical Review A}\ }\textbf {\bibinfo
  {volume} {98}},\ \bibinfo {pages} {063815} (\bibinfo {year}
  {2018})}\BibitemShut {NoStop}%
\bibitem [{\citenamefont {Schrieffer}\ and\ \citenamefont
  {Wolff}(1966)}]{schrieffer1966relation}%
  \BibitemOpen
  \bibfield  {author} {\bibinfo {author} {\bibfnamefont {J.~R.}\ \bibnamefont
  {Schrieffer}}\ and\ \bibinfo {author} {\bibfnamefont {P.~A.}\ \bibnamefont
  {Wolff}},\ }\href@noop {} {\bibfield  {journal} {\bibinfo  {journal}
  {Physical Review}\ }\textbf {\bibinfo {volume} {149}},\ \bibinfo {pages}
  {491} (\bibinfo {year} {1966})}\BibitemShut {NoStop}%
\bibitem [{\citenamefont {Davis}\ and\ \citenamefont
  {Pechukas}(1976)}]{DavisPechukas1976}%
  \BibitemOpen
  \bibfield  {author} {\bibinfo {author} {\bibfnamefont {J.~P.}\ \bibnamefont
  {Davis}}\ and\ \bibinfo {author} {\bibfnamefont {P.}~\bibnamefont
  {Pechukas}},\ }\href@noop {} {\bibfield  {journal} {\bibinfo  {journal} {The
  Journal of Chemical Physics}\ }\textbf {\bibinfo {volume} {64}},\ \bibinfo
  {pages} {3129} (\bibinfo {year} {1976})}\BibitemShut {NoStop}%
\bibitem [{\citenamefont {Wiebe}\ and\ \citenamefont
  {Babcock}(2012)}]{wiebe2012improved}%
  \BibitemOpen
  \bibfield  {author} {\bibinfo {author} {\bibfnamefont {N.}~\bibnamefont
  {Wiebe}}\ and\ \bibinfo {author} {\bibfnamefont {N.~S.}\ \bibnamefont
  {Babcock}},\ }\href@noop {} {\bibfield  {journal} {\bibinfo  {journal} {New
  Journal of Physics}\ }\textbf {\bibinfo {volume} {14}},\ \bibinfo {pages}
  {013024} (\bibinfo {year} {2012})}\BibitemShut {NoStop}%
\end{thebibliography}%

\end{appendix}
\end{document}